\documentclass[aps,prc,superscriptaddress,twoside,twocolumn,nofootinbib,10pt,%
showpacs,floatfix]{revtex4-1}

\usepackage{amsmath,amssymb}
\usepackage{graphicx,bm}
\usepackage{slashed}
\usepackage{epstopdf}
\usepackage{ulem} 
\usepackage[usenames]{color}
\usepackage{subfigure}
\usepackage{float}

\newcommand{\Slash}[1]{{\ooalign{\hfil/\hfil\crcr$#1$}}}

\allowdisplaybreaks

\begin{document}

\title{Spectral functions for $\bar{D}$ meson and $\bar{D}_{0}^{\ast}$ meson in nuclear matter with partial restoration of chiral symmetry}

\author{Daiki Suenaga}
\email{suenaga@hken.phys.nagoya-u.ac.jp}
\affiliation{Department of Physics,  Nagoya University, Nagoya, 464-8602, Japan}

\author{Shigehiro Yasui}
\email{yasuis@th.phys.titech.ac.jp}
\affiliation{Department of Physics, Tokyo Institute of Technology, Tokyo 152-8551, Japan}

\author{Masayasu Harada}
\email{harada@hken.phys.nagoya-u.ac.jp}
\affiliation{Department of Physics,  Nagoya University, Nagoya, 464-8602, Japan}

\date{\today}

\newcommand\sect[1]{\emph{#1}---}
\begin{abstract}
We investigate the in-medium masses of a $\bar{D}$ $(0^-)$ meson and a $\bar{D}_0^*$ $(0^+)$ meson and spectral functions for $\bar{D}$ 
 and $\bar{D}_0^*$ 
meson channels in nuclear matter.
These mesons are introduced as chiral partner in the chiral symmetry broken vacuum,
hence they are useful to explore the partial restoration of the broken chiral symmetry in nuclear matter.
We consider the linear sigma model to describe the chiral symmetry breaking.
Our study 
shows that the loop corrections to $\bar{D}$ and $\bar{D}_0^*$ meson masses 
provide a smaller mass splitting at finite density than that in vacuum, whose result indicates a tendency of the restoration of the chiral symmetry.
We investigate also the spectral function for $\bar{D}_0^*$ meson channel, and find
  three peaks.
The first peak which corresponds to the resonance of $\bar{D}_0^*$ meson is broadened by collisions with nucleons in medium, and the peak position shifts to 
lower mass due to the partial restoration of chiral symmetry as the density increases. 
The second peak
is identified as a threshold enhancement which shows a remarkable enhancement as the density increase. The third peak is Landau damping. 
The obtained properties of $\bar{D}$ and $\bar{D}_0^*$ mesons in nuclear matter will provide useful information for experiments.
\end{abstract}
\maketitle

\section{Introduction}

Chiral symmetry plays an important role 
not only in hadron masses but also in decay properties and inter-hadron interactions in vacuum.
Recently, to investigate the hadrons at finite density and/or temperature is an important subject, because the broken chiral symmetry can be restored partially. 
In this case, 
the linear representation is adopted, where the hadron fields are given by the irreducible representations of $SU(N_{F})_{R} \times SU(N_{F})_{L}$.

The first example is the linear sigma model by Gell-Mann and Levy~\cite{GellMann:1960np,Schwinger:1957em}.
The nucleon field is transformed linearly in the chiral group of $SU(2)_{R} \times SU(2)_{L}$.
The linear representation scheme is considered in the chiral doublet nucleons including the positive-parity state (ground state) and the negative-parity state (excited state)~\cite{Detar:1988kn}, and is applied to investigate the properties of excited baryons~\cite{Nemoto:1998um,Jido:1998av,Jido:1999hd,Jido:2001nt}.
Some studies on scalar mode ($\bar{q}q$) and pseudo-scalar mode ($\bar{q}i\gamma_5 q$) within the Nambu-Jona-Lasinio (NJL) model at density and/or temperature also exists (see Ref.~\cite{Hatsuda:1994pi} for a review and references therein).

The finite mass splitting between two states in a chiral partner is the sufficient condition for the chiral symmetry breaking in vacuum.
Therefore, to investigate the properties of chiral doublet provides us with useful information about the origin of hadron mass~(see Ref.~\cite{Hayano:2008vn} for a review and references therein).
However, the representation of the chiral group for baryons is complex, 
because they are composed at least of three quarks, $qqq$.
In this respect, heavy-light mesons, whose quark content is $\bar{Q}q$, are simpler objects, because the heavy quark is irrelevant to the chiral dynamics and the light quark component belongs to the fundamental representation of the chiral group.
In the present work, therefore, we consider heavy-light mesons as probes to explore the chiral symmetry in nuclear matter.

A heavy-light meson such as a $\bar{D}$ meson, which is composed of an anti-charm quark and a light quark, can be a good probe to investigate the chiral symmetry at finite density. Thanks to the heavy mass of the anti-charm quark, the meson mass is much larger than the low energy scale of QCD, $\Lambda_{\rm QCD}$ a few hundred MeV.
It allows us to employ a scheme of $1/m_Q$ expansion with $m_{Q}$ being the heavy quark mass, and to utilize the heavy quark spin symmetry in which the light quark spin and the heavy quark spin are almost independently conserved quantities~\cite{HQSS}.
The heavy quark symmetry is important not only in vacuum but also in nuclear matter, because it induces the channel coupling effect between a $\bar{D}$ meson and a $\bar{D}^{\ast}$ meson in interaction process with a nucleon~\cite{Yasui:2009bz,Gamermann:2010zz,GarciaRecio:2011xt,Yasui:2012rw,Suenaga:2014dia}.

On the other hand, the light quark component in a $\bar{D}$ meson shares the same light ($u$ or $d$) quark with a nucleon through the quark exchange, and can strongly interact with nuclear matter. 
Hence, the interaction between a $\bar{D}$ meson and a nucleon will be supplied mainly by the light quark dynamics, where chiral symmetry should play the significant role.
This is the reason why we focus on the heavy-light meson in nuclear matter.
Importantly, $\bar{D}$ mesons do not contain a light antiquark so that there is no 
need to take into account 
 pair annihilations of a light quark $q$ and a light antiquark $\bar{q}$.
 This is sharply in contrast to the case of a $D$ meson with the quark content $Q\bar{q}$, where the annihilation by $q\bar{q}$ and several coupled-channel processes (e.g. $DN \rightarrow \pi \Lambda_{c}, \pi \Sigma_{c}$), in which the interaction process is much complex, may provide non-negligible contributions. 
Thanks to these advantages, we can extract the informations of chiral symmetry in nuclear matter in a relatively easy way.
The experiment of heavy-light mesons at density can be performed at J-PARC and FAIR, and so on.

Several studies on heavy-light mesons at finite temperature and/or density are found in the literatures~\cite{Tsushima:1998ru,Hayashigaki:2000es,Mishra:2003se,Lutz:2005vx,Hilger:2008jg,Tolos:2009nn,Blaschke:2011yv,Yasui:2012rw,Sasaki:2014asa,Suzuki:2015est,Hattori:2015hka} (see also Ref.~\cite{Hosaka:2016ypm} for a review and references therein).
Especially in relation to chiral symmetry, in Ref.~\cite{Suenaga:2014sga}, the masses of $\bar{D}$ mesons in a dense matter such as the Skyrmion crystal was studied by including both chiral partner structure and heavy quark spin symmetry.
The chiral partners such as a $\bar{D}$ meson $(J^P=0^-)$ and a $\bar{D}_0^*$ meson $(J^P=0^+)$, whose parities are opposite each other, have unequal masses whose difference stems from the spontaneous breakdown of chiral symmetry~\cite{Nowak:1992um,Bardeen:1993ae}.
Hence, the two masses of chiral partners should be degenerate in the chiral symmetric phase.
In fact, it was found in Ref.~\cite{Suenaga:2014sga} that their masses are degenerate at certain high density; the $\bar{D}$ meson mass increases, while the $\bar{D}_0^*$ meson mass decreases, and eventually their masses become the value which is given by the averaged mass of chiral partners in the vacuum.

In Ref.~\cite{Suenaga:2014sga}, only $\sigma$ type mean field were included.
Later in Ref.~\cite{Harada:2016uca}, $\omega$ type mean field were also considered in addition, and it was found that the $\omega$ mean field affects the masses of a $D$ meson and a $\bar{D}$ meson in opposite way, while the $\omega$ mean field affects the masses of the chiral partner, a $\bar{D}$ meson and a $\bar{D}_{0}^{\ast}$ meson, in the same way.
We notice, however, that the degenerate masses of chiral partners can differ from the 
 chiral invariant mass whose value is defined by the averaged mass of chiral partners in vacuum.
We also note that, in the studies in Refs.~\cite{Suenaga:2014sga,Harada:2016uca},
only the mean fields of a $\sigma$ meson and a $\omega$ meson are taken into account, and the loop corrections from fluctuations were neglected. 

The purpose in this paper is that, based on the linear sigma model, we include, not only the mean field of a $\sigma$ meson, but also 
 loop effects perturbatively beyond the mean field level, and study the masses and the spectral functions for $\bar{D}$ and $\bar{D}_0^*$ mesons in nuclear matter.
 In our calculation,
the medium effects at one-loop order 
are mainly induced by a virtual pion and a virtual $\sigma$ meson, whose contributions are given by the mediated diagrams.
We present that the loop corrections to the mean field of a $\sigma$ meson are important for the modifications of $\bar{D}$ and $\bar{D}_0^*$ meson masses.
There are also the other diagrams 
which give the imaginary parts in self-energies 
 $\bar{D}$ and $\bar{D}_0^*$ mesons.
We will present thus that the loop corrections play the significant roles for the spectral functions for $\bar{D}$ and $\bar{D}_0^*$ mesons in nuclear matter.
The corresponding diagrams will be given concretely in the main text.
In the following, we refer ``$\bar{D}$ mesons'' as $\bar{D}$ $(J^P=0^-)$, $\bar{D}^*$ $(J^P=1^-)$, $\bar{D}_0^*$ $(J^P=0^+)$ and $\bar{D}_1$ $(J^P=1^+)$ mesons for short notations, but this will not cause any confusions. 

This paper is organized as follows.
In Sec.~\ref{sec:LinearSigma}, we give a scheme for describing a nuclear matter by using the linear sigma model, and consider the pion and $\sigma$ meson fluctuations to satisfy fully the chiral symmetry at finite density. 
In Sec.~\ref{sec:DLagrangian}, we introduce the ``$\bar{D}$ mesons'' as chiral partners, and formulate the Lagrangian satisfying both chiral symmetry and heavy quark spin symmetry.
In Sec.~\ref{sec:DInMedium}, we study the mass and the spectral function of $\bar{D}$ and $\bar{D}_0^*$ meson in nuclear matter.
In Sec~\ref{sec:Summary}, we give a conclusion and discussions.

\section{Construction of nuclear matter}
\label{sec:LinearSigma}
\subsection{Lagrangian and gap equation}
In the linear sigma model,
 the nuclear matter effect for ``$\bar{D}$ mesons'' are mediated by pion and $\sigma$ mesons interacting with nucleons in nuclear matter.
Before investigating the ``$\bar{D}$ mesons'', first of all, we need to give nuclear matter by using 
the degrees of freedom of a pion, a $\sigma$ meson and a nucleon in the linear sigma model.
We assume the spatially uniform and isospin-symmetric nuclear matter.
Especially, we need the spectral functions for pion and $\sigma$ meson in nuclear matter to compute the modifications of ``$\bar{D}$ mesons'' in medium, as it will be discussed in the later section.

We employ the linear sigma model which is invariant under the $SU(2)_R \times SU(2)_L$ chiral symmetry~\cite{GellMann:1960np} with finite baryon number chemical potential $\mu_B$~\cite{Birse:1994cz}:
\begin{eqnarray}
{\cal L}_{\rm LS} &=& \bar{\psi}(i\Slash{\partial}+\mu_B\gamma^0)\psi-g\bar{\psi}(\sigma+i\gamma_5\tau^a\pi^a)\psi \nonumber\\
&&+\frac{1}{2}\partial_{\mu}\sigma\partial^{\mu}\sigma+\frac{1}{2}\partial_{\mu}\pi^a\partial^{\mu}\pi^a-\frac{m_0^2}{2}(\sigma^2+\pi^2) \nonumber\\
&&-\frac{\lambda}{4}(\sigma^2+\pi^2)^2+\epsilon\sigma
\label{LinearSigma} \ ,
\end{eqnarray}
where the explicit breaking term ${\cal L}_{\rm ex}=\epsilon\sigma$ is added to reproduce pion pass $m_\pi=138$ MeV, and $\sigma$ and $\pi^a$ stand for the $\sigma$ meson field and the pion field, respectively. $\psi$ is the doublet of nucleon fields and $\bar{\psi}$ is defined as $\bar{\psi}=\psi^{\dagger}\gamma^0$, and $\tau^a$ ($a=1,2,3$) is the Pauli matrix for isospin.
We adopt $g$, $m_0^2<0$ and $\lambda>0$ as free parameters, 
whose values are fitted to reproduce the properties of 
vacuum, and are kept as constant numbers from vacuum to finite density.

\begin{figure}[thbp]
\centering
\includegraphics*[scale=0.7]{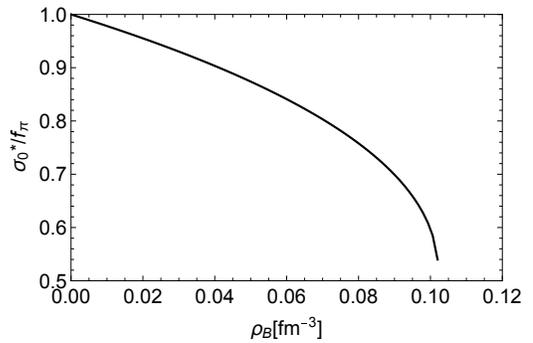}
\caption{Density dependence of the mean field $\sigma_0^*$.}
\label{fig:MeanValue}
\end{figure}

The effective action for pion and $\sigma$ meson is given by performing the path integral for $\psi$ and $\bar{\psi}$ at nucleon one-loop order from the Lagrangian~(\ref{LinearSigma}).
The obtained effective potential $V[\sigma,\pi]$ reads 
\begin{eqnarray}
V[\sigma_0^*,0] &=& 2i 
\int\frac{\tilde{d}^4k}{(2\pi)^4}\, {\rm ln}\left(\Slash{k}+\mu_B\gamma^0-g\sigma_0^*\right) \nonumber\\
&&+\frac{m_0^2}{2}\sigma_0^{*2}+\frac{\lambda}{4}\sigma_0^{*4}+\epsilon\sigma_0^*, \label{EffectivePotential}
\end{eqnarray}
for the mean field of the $\sigma$ field at finite density, $\sigma_{0}^{\ast}$, 
where the mean field of a pion field is set to be zero, as we have assumed the parity conservation in the ground state. 
``$\int\tilde{d}^4k/(2\pi)^4$'' in Eq.~(\ref{EffectivePotential}) stands for the momentum integral which is dependent on Fermi momentum $k_F$ explicitly. Namely, when we define an integral $I_{k_F}$ of a function $f(k_0,\vec{k};k_F)$ by
\begin{eqnarray}
I_{k_F} &\equiv& \int\frac{d^4k}{(2\pi)^4}f(k_0,\vec{k};k_F) \ ,
\end{eqnarray}
``$\int\tilde{d}^4k/(2\pi)^4$'' integral refers
\begin{eqnarray}
\int\frac{\tilde{d}^4k}{(2\pi)^4}f(k_0,\vec{k};k_F) \equiv  I_{k_F}-I_{k_F=0}\ . \label{TildeInt}
\end{eqnarray}
The Fermi momentum $k_F$ is defined by the relation $\sqrt{k_F^2+m_N^{*2}}=\mu_B$ with $m_{N}^{\ast}=g\sigma_{0}^{\ast}$ being the nucleon mass at finite density. 
The magnitude of the chiral symmetry breaking is controlled by $\sigma_{0}^{\ast}$, where the finite density effect is accounted by $k_{F}$ via the nucleon number density $\rho_B = \frac{2k_F^3}{3\pi^2}$.
In the present study, we focus on the contribution of nucleons in the Fermi surface, and neglect the antinucleons in vacuum.

 The mean field $\sigma_0^*$ is determined by the stationary point of $V[\sigma_0^*,0]$ respect to $\sigma_0^*$: $\frac{\partial}{\partial \sigma_0^*}V[\sigma_0^*,0]=0$.
 It leads to the gap equation:
\begin{eqnarray}
m_0^2+\lambda\sigma_0^{*2}-\frac{\epsilon}{\sigma_0^*} = -\frac{ 2 g  }{\pi^2}\int_0^{k_F}d|\vec{k}|\frac{|\vec{k}|^2}{\sqrt{|\vec{k}|^2+m_N^{*2}}}\ .
\label{GapDensDep}
\end{eqnarray}
We notice that, the mean field $\sigma_0^*$ at zero density is identical to the pion decay constant $f_\pi$ in vacuum, and $f_\pi$ satisfies the gap equation~(\ref{GapDensDep}) with $k_F=0$, which is given in the third line in Eq.~(\ref{Parameters}) below. 

The gap equation~(\ref{GapDensDep}) determines the density dependence of the mean field $\sigma_0^*$ in nuclear matter. The parameters  $g$, $m_0^2$, $\lambda$ and $\epsilon$ are determined by the physical quantities in vacuum: nucleon mass $m_N=939$ MeV, sigma term $\Sigma_{\pi N} = 45$ MeV, pion decay constant $f_\pi=92.4$ MeV and pion mass $m_\pi=138$ MeV.
They are determined via the following relations~\footnote{The sigma term 
 is defined as $\Sigma_{\pi N} \equiv -f_\pi m_\pi^2\left.\frac{\partial\sigma_0^*}{\partial \rho_B}\right|_{\rho_B=0}$, and it is obtained by taking the derivative of $\rho_B$ for the both side of Eq.~(\ref{GapDensDep}) where $\rho_B = \frac{2k_F^3}{3\pi^2}$ is used. }:
\begin{eqnarray}
m_N&=& gf_\pi, \nonumber\\
\Sigma_{\pi N} &=& g  \frac{f_\pi^2m_\pi^2}{2\lambda f_\pi^3+\epsilon},  \nonumber\\
m_0^2+\lambda f_\pi^2&=&\frac{\epsilon}{f_\pi}, \nonumber\\
m_\pi^2 &=& \frac{\epsilon}{f_\pi}. \label{Parameters}
\end{eqnarray}
We obtain 
\begin{eqnarray}
g &=& 10.2, \nonumber\\
\lambda &=& 22.2, \nonumber\\
m_0^2 &=& -1.70 \times 10^5\ [{\rm MeV}^2],  \nonumber\\
\epsilon &=& 1.76\times 10^6 \ [{\rm MeV}^3].
\end{eqnarray}
We plot the density dependence of the mean field $\sigma_0^*$ in Fig.~\ref{fig:MeanValue}.
We confirm that $\sigma_0^*$ decreases as the density increases, indicating the partial restoration of chiral symmetry~\cite{Birse:1994cz}.

\begin{figure*}[htbp]
\centering
\includegraphics*[scale=0.55]{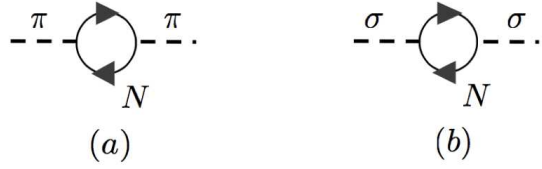}
\caption{Self-energies of (a) the pion and (b) the $\sigma$ meson.  The nucleon loop is calculated by using in-medium propagator $\tilde{G}_N(k_0,\vec{k})$ in Eq.~(\ref{InMediumPropagator}).}
\label{fig:SelfEnergies}
\end{figure*}

\begin{figure*}[htbp]
\centering
\includegraphics*[scale=0.5]{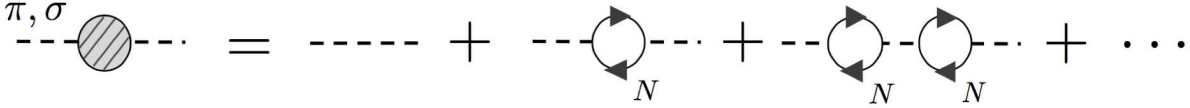}
\caption{Resummed propagator for pion and $\sigma$ meson.}
\label{fig:ResummedPropagator}
\end{figure*}


\begin{figure*}[t]
  \begin{center}
    \begin{tabular}{c}

      \begin{minipage}{0.5\hsize}
        \begin{center}
         \includegraphics*[scale=0.7]{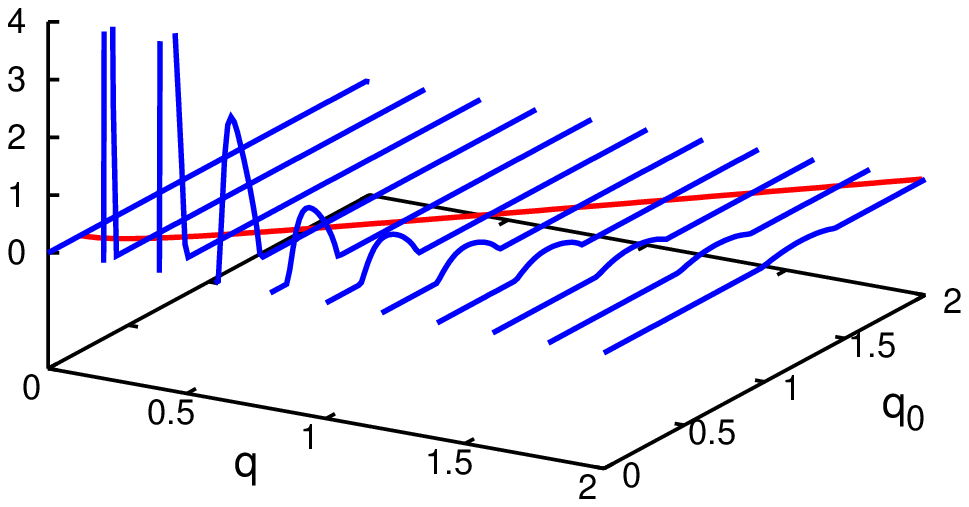}\\
          \hspace{0.7cm} (a) Spectral function for pion $\rho_\pi^*(q_0,\vec{q})$
        \end{center}
      \end{minipage}

      \begin{minipage}{0.5\hsize}
        \begin{center}
          \includegraphics*[scale=0.7]{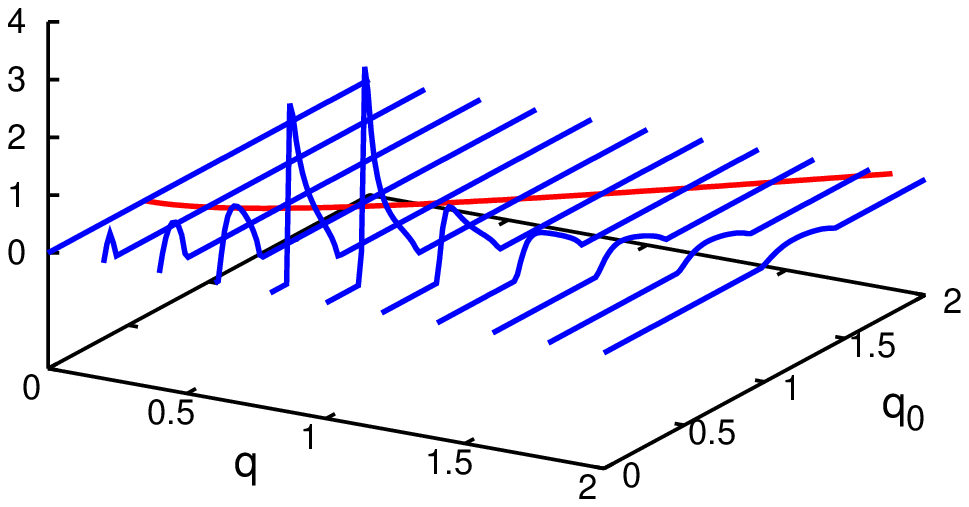}\\
          \hspace{0.7cm} (b) Spectral function for $\sigma$ meson $\rho_\sigma^*(q_0,\vec{q})$
        \end{center}
      \end{minipage}

      \end{tabular}
 \caption{(color online) Spectral functions for (a) pion $\rho_\pi^*(q_0,\vec{q})$ and (b) $\sigma$ meson $\rho_\sigma^*(q_0,\vec{q})$ at $\rho_B=0.066$ [fm$^{-3}$]. $q$ is defined by $q=|\vec{q}|$. $q$, $q_0$ and $\rho_{\pi(\sigma)}^*(q_0,\vec{q})$ in these figures are scaled by $m_N^*$ so that these are dimensionless quantities. Red curves indicate the dispersion relations of (a) pion and (b) $\sigma$ meson, which are determined by $\tilde{\Gamma}_{\pi}^{*(2)} (q_0,\vec{q}) = 0$ or $\tilde{\Gamma}_{\sigma}^{*(2)} (q_0,\vec{q}) = 0$, respectively. Then red curves should be understood as the position of the delta function which refers the one particle state of pion and $\sigma$ meson. Bumps in the space-like region are induced by the Landau damping which is diagrammatically shown in Fig.~\ref{fig:LandauBump}}
\label{fig:Spectrals}
  \end{center}
\end{figure*}
\subsection{Fluctuation of $\sigma$ meson and pion in the nuclear matter}
\label{sec:FermionOneLoop}

For later use,
we need to calculate the spectral functions for pion and $\sigma$ meson in nuclear matter.
They will be used to compute the medium modifications to ``$\bar{D}$ mesons'', as demonstrated in Sec.~\ref{sec:DInMedium}. 
To obtain the spectral function, we calculate the two-point functions for pion and $\sigma$ meson by using the effective action (\ref{EffectivePotential}), provided that the mean field $\sigma_{0}^{\ast}$ is replaced to a sum of the background (mean field) $\sigma_0^*$ and the fluctuation part $\sigma$ (this $\sigma$ should not be confused with the original field):
\begin{widetext}
\begin{eqnarray}
\Gamma[\sigma,\pi] &=&  -i{{\rm Tr}}\, {\rm ln}(i\Slash{\partial}+\mu_B\gamma^0-g(\sigma_0^*+\sigma+i\gamma_5\tau^a\pi^a))\nonumber\\
&& +\int d^4x\left(\frac{1}{2}\partial_{\mu}\sigma\partial^{\mu}\sigma+\frac{1}{2}\partial_{\mu}\pi^a\partial^{\mu}\pi^a-\frac{m_0^2}{2}((\sigma_0^*+\sigma)^2+\pi^2)-\frac{\lambda}{4}((\sigma_0^*+\sigma)^2+\pi^2)^2+\epsilon(\sigma_0^*+\sigma)\right) \nonumber\\ \nonumber\\
&=& -i{{\rm Tr}}\, {\rm ln}(i\Slash{\partial}+\mu_B\gamma^0)+i\frac{g^2}{2}{{\rm Tr}}\left[\frac{1}{i\Slash{\partial}+\mu_B\gamma^0}\sigma\frac{1}{i\Slash{\partial}+\mu_B\gamma^0}\sigma\right] +i\frac{g^2}{2} {\rm Tr}\left[\frac{1}{i\Slash{\partial}+\mu_B\gamma^0}i\gamma_5\pi^a\tau^a\frac{1}{i\Slash{\partial}+\mu_B\gamma^0}i\gamma_5\pi^b\tau^b\right] \nonumber\\
&& + \int d^4x\left\{ \frac{1}{2}\partial_\mu\sigma\partial^\mu\sigma+\frac{1}{2}\partial_\mu \pi^a\partial^\mu\pi^a-\frac{1}{2}(m_0^2+3\lambda\sigma_0^{*2})\sigma^2-\frac{1}{2}(m_0^2+\lambda\sigma_0^{*2})\pi^a\pi^a \right\}+ ({\rm interactions}  )
\ . 
\label{EffectiveAction}
\end{eqnarray}
\end{widetext}
In obtaining the second equality in Eq.~(\ref{EffectiveAction}), we have expanded the logarithmic function with respect to $\sigma$ and $\pi^{a}$, and have used the gap equation~(\ref{GapDensDep}) to eliminate 
 the tadpole diagrams.
 Higher order terms of $\sigma$ meson and pion fields are abbreviated 
  in ``$({\rm interactions})$''. 
  
The two-point vertex functions for $\sigma$ and $\pi$ mesons $\tilde{\Gamma}^{*(2)}_{\sigma}(q_0,\vec{q})$ and $\tilde{\Gamma}^{*(2)}_\pi(q_0,\vec{q})$ ($\tilde{\Gamma}^{*(2)}_{\pi, ab}(q_0,\vec{q})=\delta^{ab}\tilde{\Gamma}^{*(2)}_\pi(q_0,\vec{q})$ with isospin indices $a=1,2,3$), each of which is defined as the inverse of propagator with momentum $(q_0,\vec{q})$, are given as 
\begin{widetext}
\begin{eqnarray}
\tilde{\Gamma}^{*(2)}_\pi(q_0,\vec{q})&=&q^2-(m_0^2+\lambda\sigma_0^{*2})- 2i g^2  \int\frac{\tilde{d}^4k}{(2\pi)^4}{\rm tr}\left[i\gamma_5\tilde{G}_N(k_0,\vec{k})i\gamma_5\tilde{G}_N(k_0-q_0,\vec{k}-\vec{q})\right]\nonumber\\
&\equiv&q^2-m_\pi^{*2}-i\tilde{\Sigma}^{*}_\pi(q_0,\vec{q}) \label{TPPi} \ ,
\end{eqnarray}
and
\begin{eqnarray}
\tilde{\Gamma}^{*(2)}_\sigma(q_0,\vec{q})&=&q^2-(m_0^2+3\lambda\sigma_0^{*2})-2i g^2 \int\frac{\tilde{d}^4k}{(2\pi)^4}{\rm tr}\left[\tilde{G}_N(k_0,\vec{k})\tilde{G}_N(k_0-q_0,\vec{k}-\vec{q})\right] \nonumber\\
&\equiv&q^2-m_\sigma^{*2}-i\tilde{\Sigma}^{*}_\sigma(q_0,\vec{q})  \label{TPSigma}\ ,
\end{eqnarray}
\end{widetext}
where $m_\pi^{*2}=m_0^2+\lambda\sigma_0^{*2}$ and $m_\sigma^{*2}=m_0^2+3\lambda\sigma_0^{*2}$ are the ``bare" mass of pion and $\sigma$ meson excluding the nucleon-hole effect.
The (momentum-dependent) in-medium masses are given by $m_\pi^{*2}+i\tilde{\Sigma}^{*}_\pi(q_0,\vec{q})$ and $m_\sigma^{*2}+i\tilde{\Sigma}^{*}_\sigma(q_0,\vec{q})$ for each. 
Again, ``$\int \tilde{d}^4k/(2\pi)^4$'' stands for the momentum integration dependent on Fermi momentum $k_F$, as introduced in Eq.~(\ref{TildeInt}), to be consistent with the gap equation~(\ref{GapDensDep}).
 $\tilde{G}_N(k_0,\vec{k})$ is the in-medium propagator of nucleon with momentum $k^\mu = (k_0,\vec{k})$~\cite{TFT}:
\begin{eqnarray}
\tilde{G}_N(k_0,\vec{k}) &=& (\Slash{k}+m_N^{*})
\biggl[ \frac{i}{k^2-m_N^{*2}+i\epsilon} 
\nonumber\\
&& 
-2\pi\theta(k_0)\theta(k_F-|\vec{k}|)\delta(k^2-m_N^{*2})
\biggr] \ ,
 \label{InMediumPropagator}
\end{eqnarray}
with an infinitely small positive number $\epsilon>0$.
The self-energies $\tilde{\Sigma}_\pi^*(q_0,\vec{q})$ and $\tilde{\Sigma}_\sigma^*(q_0,\vec{q})$ are diagrammatically shown in Fig.~\ref{fig:SelfEnergies}. The detailed calculation of these self-energies are performed in Appendix.~\ref{sec:OneLoopApp}. The inverse of two-point vertex function, $\tilde{\Gamma}^{*(2)}_\pi(q_0,\vec{q})$ and $\tilde{\Gamma}_\sigma^{*(2)}(q_0,\vec{q})$, provides the corresponding propagator of pion or $\sigma$ meson in nuclear matter, as infinite sums of $\tilde{\Sigma}_\pi^*(q_0,\vec{q})$ and $\tilde{\Sigma}_\sigma^*(q_0,\vec{q})$ in Fig.~\ref{fig:ResummedPropagator}.
We thus need to utilize those resummed propagators to fulfill the chiral symmetry for the fluctuations of pion and $\sigma$ meson in nuclear matter, as explained in Appendix~\ref{sec:OneLoopApp}.

\begin{figure}[htbp]
\centering
\includegraphics*[scale=0.5]{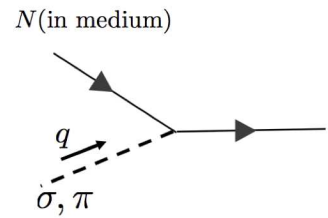}
\caption{Diagrammatical interpretation for the Landau damping. This is understood as the scattering process between a pion (a $\sigma$ meson) and a nucleon in medium since four-momentum of pion ($\sigma$ meson) is in the space-like domain $q^2<0$. }
\label{fig:LandauBump}
\end{figure}

The resulting in-medium spectral functions for pion and $\sigma$ meson are given by  
\begin{eqnarray}
&&\rho^*_{\pi(\sigma)}(q_0,\vec{q}) \nonumber\\
&=& \frac{-2{\rm Im}\tilde{\Sigma}_{\pi(\sigma)}^{*R}(q_0,\vec{q})}{\left[q^2-m_{\pi(\sigma)}^{*2}-{\rm Re}\tilde{\Sigma}_{\pi(\sigma)}^{*R}(q_0,\vec{q})\right]^2+\left[{\rm Im}\tilde{\Sigma}_{\pi(\sigma)}^{*R}(q_0,\vec{q})\right]^2},  \nonumber\\
\label{RenormalizedSpectral}
\end{eqnarray}
where the retarded self-energy $\tilde{\Sigma}_{\pi(\sigma)}^{*R}(q_0,\vec{q})$ is related to the self-energy 
$\tilde{\Sigma}^*_{\pi(\sigma)}(q_0,\vec{q})$ in Eq.~(\ref{TPPi}) as
\begin{eqnarray}
{\rm Re}\tilde{\Sigma}_{\pi(\sigma)}^{*R}(q_0,\vec{q})={\rm Re}\left(i\tilde{\Sigma}^*_{\pi(\sigma)}(q_0,\vec{q})\right)\ , \label{RelationRe}
\end{eqnarray}
and
\begin{eqnarray}
{\rm Im}\tilde{\Sigma}_{\pi(\sigma)}^{*R}(q_0,\vec{q}) =\epsilon(q_0){\rm Im}\left(i\tilde{\Sigma}_{\pi(\sigma)}^*(q_0,\vec{q})\right)\ . \label{RelationIm}
\end{eqnarray}
$\epsilon(q_0)$ is the sign function defined as $\epsilon(q_0) = +1(-1)$ for $q_0>0(q_0<0)$. 
The detailed derivation of the spectral function is given in Appendix.~\ref{sec:Dispersion}. The dependence on energy $q_{0}$ and momentum $q \equiv \vert \vec{q} \vert$ at $\rho_B=0.066$ [fm$^{-3}$] is
 displayed in Fig.~\ref{fig:Spectrals} as an example. Bumps in the space-like region corresponds to the Landau damping, i.e., the space-like ($q^2<0$) pion ($\sigma$ meson) is absorbed by a nucleon in medium 
which is diagrammatically shown in Fig.~\ref{fig:LandauBump}.
 
The red curve in Fig.~\ref{fig:Spectrals} (a) and (b) shows the dispersion relation for one particle state of pion and $\sigma$ meson, which are defined as the solution of
\begin{eqnarray}
\Gamma_{\pi}^{*(2)} (q_0,\vec{q}) &=& 0 \ ,\nonumber\\
\Gamma_{\sigma}^{*(2)} (q_0,\vec{q}) &=& 0\ , \label{OPDispersion}
\end{eqnarray}
respectively. Note that the solution of Eq.~(\ref{OPDispersion}) does not include any imaginary parts, so that the red curve in Fig.~\ref{fig:Spectrals} (a) and (b) should be understood as the position of the delta function which refers the one particle state of pion or $\sigma$ meson. 

Utilizing the spectral functions for pion and $\sigma$ meson in Eq.~(\ref{RenormalizedSpectral}), we shall compute the modifications of ``$\bar{D}$ mesons'' in nuclear matter in Sec.~\ref{sec:DInMedium}.
Before going to this discussion, in the next section, we give a formulation for the interaction between a pion (a $\sigma$ meson) and ``$\bar{D}$ mesons''.

 
\section{Lagrangian for ``$\bar{D}$ mesons'' with chiral partner structure}
\label{sec:DLagrangian}
In this section, we introduce the Lagrangian for ``$\bar{D}$ mesons'' which is based on the idea of chiral partner structure~\cite{Nowak:1992um,Bardeen:1993ae}
 and heavy quark spin symmetry~\cite{HQSS}.
The mass difference between ($\bar{D},\bar{D}^*$) and ($\bar{D}_0^*,\bar{D}_1$) is the finite value in vacuum because it is induced by the spontaneous breakdown of chiral symmetry.
As the property in the heavy quark spin symmetry, the masses of ($\bar{D},\bar{D}^*)$ 
 mesons can be regarded to be the same, and those of ($\bar{D}^*_0,\bar{D}_1)$ 
 mesons can also be regarded to be the same, as each mass splitting is suppressed by the factor $1/m_{Q}$.

The Lagrangian of ``$\bar{D}$ mesons'', in which interactions with a $\sigma$ meson and a pion are introduced based on the chiral partner structure, is invariant under the $SU(2)_L \times SU(2)_R$ chiral symmetry as well as under $SU(2)_S$ heavy quark spin symmetry~\cite{Nowak:1992um,Bardeen:1993ae}:
\begin{eqnarray}
{\cal L}_{\mathrm{HMET}} &=& {\rm tr}[H_L(iv\cdot\partial)\bar{H}_L]+{\rm tr}[H_R(iv\cdot\partial)\bar{H}_R] \nonumber\\
&& +\frac{\Delta_m}{2f_{\pi}}{\rm tr}[H_LM\bar{H}_R \!+\! H_RM^{\dagger}\bar{H}_L] \nonumber\\
&& +i\frac{g_A}{2f_{\pi}}{\rm tr}[H_R\gamma_5\gamma^{\mu}\partial_{\mu}M^{\dagger}\bar{H}_L \!-\! H_L\gamma_5\gamma^{\mu}\partial_{\mu}M\bar{H}_R], 
\nonumber\\
 \label{HMETLagrangian}
\end{eqnarray}
where $v^{\mu}$ is the four-velocity of reference of $H_L$ and $H_R$ fields, and $\Delta_m$ is identified as the mass difference between chiral partners,  as will be seen later~\footnote{As derived in Appendix.A in Ref.~\cite{Suenaga:2015daa}, other interaction terms up to $O(\partial^2M)$ can show up in addition to third line in Eq.~(\ref{HMETLagrangian}), however, we assume that this term is dominant compared with the other terms as done in the main parts of analysis in Ref.~\cite{Suenaga:2015daa}.}.
$M$ is the chiral field defined by $M=\sigma+i\pi^a\tau^a$ which transforms under the $SU(2)_L\times SU(2)_R$ chiral transformation as
\begin{eqnarray}
M \to g_L M g_R^\dagger\ ,
\end{eqnarray}
where $g_L$ $(g_R)$ is the element of $SU(2)_L$ $(SU(2)_R)$. $H_L$, $H_R$ are heavy meson field which are schematically drawn as $H_{L(R)}\sim Q\bar{q}_{L(R)}$.
We define $\bar{H}_{L(R)}=\gamma^0 H_{L(R)}\gamma^0$. Then $H_L$ and $H_R$ transform under the $SU(2)_L\times SU(2)_R$ chiral transformation as
\begin{eqnarray}
H_{L(R)} &\to&  H_{L(R)}g^\dagger_{L(R)} \ ,
\end{eqnarray} 
and $SU(2)_S$ heavy quark spin transformation as
\begin{eqnarray}
H_{L(R)} &\to& S H_{L(R)} \ ,
\end{eqnarray}
where $S$ is the element of $SU(2)_S$. 
$H_L$ and $H_R$ are related to the field $H$ and $G$ whose parities are positive and negative, respectively, through the following relations:
\begin{eqnarray}
H_L &=& \frac{1}{\sqrt{2}}\left[G+iH\gamma_5\right] \ , 
\\
H_R &=& \frac{1}{\sqrt{2}}\left[G-iH\gamma_5\right] \ . \\ \nonumber
\end{eqnarray}
Then $H$ field contains odd-parity states, $D(0^-)$ and $D^*(1^-)$ mesons, while the $G$ field contains even-parity states, $D_0^*(0^+)$ and $D_1(1^+)$ mesons.
Namely, $H$ field and $G$ field can be parametrized 
as
\begin{eqnarray}
H &=& \frac{1+\Slash{v}}{2}[i\gamma_5D_v+\Slash{D}_v^*] \ , \label{PhysicalStates1} 
\\  \nonumber\\
G &=& \frac{1+\Slash{v}}{2}[D_{0v}^*-i\Slash{D}_{1v}\gamma_5]\ ,  \label{PhysicalStates2}\\ \nonumber
\end{eqnarray}
where the Lorentz index $\mu$ is adopted for the vector and axial-vector mesons, as $D^{\ast \mu}_{v}$ and $D_{1v}^{\mu}$, respectively. The subscript $v$ are added so that they refer ``$D$ mesons'' fields defined within the heavy meson effective theory.

By rewriting the Lagrangian~(\ref{HMETLagrangian}) of Eqs.~(\ref{PhysicalStates1}) and (\ref{PhysicalStates2}), 
we get the Lagrangian expressed explicitly in terms of the $D$, $D^*$, $D_0^*$ and $D_1$ meson fields. 
We notice that, in the present analysis, 
we do not consider ``$D$ mesons'' ($\sim Q\bar{q}$) but ``$\bar{D}$ mesons'' ($\sim \bar{Q}q$) in nuclear matter in order to avoid $q\bar{q}$ pair annihilation processes and other channel-coupled processes.
Hence we need the Lagrangian of ``$\bar{D}$ mesons''.
This is obviously obtained by taking the charge conjugation to ``$D$ mesons'', which leads to
\begin{widetext}
\begin{eqnarray}
{\cal L}_{\mathrm{HMET}} &=&  2\bar{D}_v(iv\cdot\partial)\bar{D}_v^{\dagger}-2\bar{D}_{v\mu}^*(iv\cdot\partial)\bar{D}_v^{*\dagger\mu} + 2\bar{D}_{0v}^*(iv\cdot\partial)\bar{D}_{0v}^{*\dagger}-2\bar{D}_{1v\mu}(iv\cdot\partial)\bar{D}_{1v}^{\dagger\mu}  \nonumber\\
&& +  \frac{\Delta_m}{2f_{\pi}}[\bar{D}_{0v}^*(M+M^{\dagger})\bar{D}_{0v}^{*\dagger}-\bar{D}_{1v\mu}(M+M^{\dagger})\bar{D}_{1v}^{\dagger\mu}-\bar{D}_v(M+M^{\dagger})\bar{D}^{\dagger}_v+\bar{D}_{v\mu}^*(M+M^{\dagger})\bar{D}_v^{*\dagger\mu}] \nonumber\\
 && +\frac{\Delta_m}{2f_{\pi}}[\bar{D}_{0v}^*(M-M^{\dagger})\bar{D}_v^{\dagger}-\bar{D}_{1v\mu}(M-M^{\dagger})\bar{D}_v^{*\dagger\mu}-\bar{D}_v(M-M^{\dagger})\bar{D}_{0v}^{*\dagger}+\bar{D}_{v\mu}^{*\dagger}(M-M^{\dagger})\bar{D}_{1v}^{\dagger\mu}]\nonumber\\ 
&& -\frac{g_A}{2f_{\pi}}[\bar{D}_{1v}^{\mu}(\partial_{\mu}M^{\dagger}-\partial_{\mu}M)\bar{D}_{0v}^{*\dagger}-\bar{D}_{0v}^{*}(\partial_{\mu}M^{\dagger}-\partial_{\mu}M)\bar{D}_{1v}^{\dagger\mu}-\epsilon^{\mu\nu\rho\sigma}\bar{D}_{1v\mu}(\partial_{\nu}M^{\dagger}-\partial_{\nu}M)\bar{D}_{1v\rho}^{\dagger}v_{\sigma}] \nonumber\\
&& +\frac{g_A}{2f_{\pi}}[\bar{D}_v^{*\mu}(\partial_{\mu}M^{\dagger}-\partial_{\mu}M)\bar{D}^{\dagger}_v-\bar{D}_v(\partial_{\mu}M^{\dagger}-\partial_{\mu}M)\bar{D}_v^{*\dagger\mu}-\epsilon^{\mu\nu\rho\sigma}\bar{D}_{v\mu}^*(\partial_{\nu}M^{\dagger}-\partial_{\nu}M)\bar{D}_{v\rho}^{*\dagger}v_{\sigma}] \nonumber\\
&& +\frac{g_A}{2f_{\pi}}[\bar{D}_{1v}^{\mu}(\partial_{\mu}M^{\dagger}+\partial_{\mu}M)\bar{D}_v^{\dagger}+\bar{D}_v(\partial_{\mu}M^{\dagger}+\partial_{\mu}M)\bar{D}_{1v}^{\dagger\mu}]\nonumber\\
& & -\frac{g_A}{2f_{\pi}}[\bar{D}_{0v}^*(\partial_{\mu}M^{\dagger}+\partial_{\mu}M)\bar{D}^{*\dagger\mu}_v+\bar{D}_v^{*\mu}(\partial_{\mu}M^{\dagger}+\partial_{\mu}M)\bar{D}_{0v}^{*\dagger}]\nonumber\\
& & -\frac{g_A}{2f_{\pi}}[\epsilon^{\mu\nu\rho\sigma}\bar{D}_{1v\nu}(\partial_{\rho}M^{\dagger}+\partial_{\rho}M)\bar{D}_{v\mu}^{*\dagger}v_{\sigma}+\epsilon^{\mu\nu\rho\sigma}\bar{D}_{v\mu}^*(\partial_{\rho}M^{\dagger}+\partial_{\rho}M)\bar{D}_{1v\nu}^{\dagger}v_{\sigma}] \ . \label{HMET}
\end{eqnarray}
This is the Lagrangian form which is valid in the heavy quark limit ($m_{Q} \rightarrow \infty$).
To study the spectral functions for ``$\bar{D}$ mesons'', 
  it is useful to further rewrite the Lagrangian in the relativistic form, keeping the finite mass of the heavy mesons, as 
\begin{eqnarray}
{\cal L} &=&   \partial_{\mu}\bar{D}_0^*\partial^{\mu}\bar{D}_0^{* \dagger}-m^2\bar{D}_0^*\bar{D}_0^{*\dagger}-\partial_{\mu}\bar{D}_{1\nu}\partial^{\mu}\bar{D}_1^{\dagger\nu}+\partial_{\mu}\bar{D}_{1\nu}\partial^{\nu}\bar{D}_1^{\dagger\mu}+m^2\bar{D}_{1\mu}\bar{D}_1^{\dagger\mu} \nonumber\\
&& +\partial_{\mu}\bar{D}\partial^{\mu}\bar{D}^{\dagger}-m^2\bar{D}\bar{D}^{\dagger}-\partial_{\mu}\bar{D}^*_{\nu}\partial^{\mu}\bar{D}^{*\dagger\nu}+\partial_{\mu}\bar{D}^*_{\nu}\partial^{\nu}\bar{D}^{*\dagger\mu}+m^2\bar{D}^*_{\mu}\bar{D}^{*\dagger\mu} \nonumber\\
&& -\frac{\Delta_m}{2f_\pi}m[\bar{D}_0^*(M+M^{\dagger})\bar{D}_0^{*\dagger}-\bar{D}_{1\mu}(M+M^{\dagger})\bar{D}_1^{\dagger\mu}-\bar{D}(M+M^{\dagger})\bar{D}^{\dagger}+\bar{D}_{\mu}^*(M+M^{\dagger})\bar{D}^{*\mu\dagger}] \nonumber\\
 && -\frac{\Delta_m}{2f_\pi}m[\bar{D}_0^*(M-M^{\dagger})\bar{D}^{\dagger}-\bar{D}_{1\mu}(M-M^{\dagger})\bar{D}^{*\dagger\mu}-\bar{D}(M-M^{\dagger})\bar{D}_0^{*\dagger}+\bar{D}_{\mu}^{* }(M-M^{\dagger})\bar{D}_1^{\dagger\mu}]\nonumber\\ 
&& -\frac{g_A}{2}\frac{m}{f_{\pi}}[\bar{D}_1^{\mu}(\partial_{\mu}M^{\dagger}-\partial_{\mu}M)\bar{D}_0^{*\dagger}-\bar{D}_0^{*}(\partial_{\mu}M^{\dagger}-\partial_{\mu}M)\bar{D}_1^{\dagger\mu}-\frac{1}{m}\epsilon^{\mu\nu\rho\sigma}\bar{D}_{1\mu}(\partial_{\nu}M^{\dagger}-\partial_{\nu}M)i\partial_{\sigma}\bar{D}_{1\rho}^{\dagger}] \nonumber\\
&& +\frac{g_A}{2}\frac{m}{f_{\pi}}[\bar{D}^{*\mu}(\partial_{\mu}M^{\dagger}-\partial_{\mu}M)\bar{D}^{\dagger}-\bar{D}(\partial_{\mu}M^{\dagger}-\partial_{\mu}M)\bar{D}^{*\dagger\mu}-\frac{1}{m}\epsilon^{\mu\nu\rho\sigma}\bar{D}_{\mu}^*(\partial_{\nu}M^{\dagger}-\partial_{\nu}M)i\partial_{\sigma}\bar{D}_{\rho}^{*\dagger}] \nonumber\\
&& +\frac{g_A}{2}\frac{m}{f_{\pi}}[\bar{D}_1^{\mu}(\partial_{\mu}M^{\dagger}+\partial_{\mu}M)\bar{D}^{\dagger}+\bar{D}(\partial_{\mu}M^{\dagger}+\partial_{\mu}M)\bar{D}_1^{\dagger\mu}]\nonumber\\
& & -\frac{g_A}{2}\frac{m}{f_{\pi}}[\bar{D}_0^*(\partial_{\mu}M^{\dagger}+\partial_{\mu}M)\bar{D}^{*\dagger\mu}+\bar{D}^{*\mu}(\partial_{\mu}M^{\dagger}+\partial_{\mu}M)\bar{D}_0^{*\dagger}]\nonumber\\
& & -\frac{g_A}{2}\frac{1}{f_{\pi}}[\epsilon^{\mu\nu\rho\sigma}\bar{D}_{1\nu}(\partial_{\rho}M^{\dagger}+\partial_{\rho}M)i\partial_{\sigma}\bar{D}_{\mu}^{*\dagger}+\epsilon^{\mu\nu\rho\sigma}\bar{D}_{\mu}^*(\partial_{\rho}M^{\dagger}+\partial_{\rho}M)i\partial_{\sigma}\bar{D}_{1\nu}^{\dagger}] \ , \label{StartingLagrangian}
\end{eqnarray}
\end{widetext}
with $m$ being the average mass of ($\bar{D},\bar{D}^*$) and ($\bar{D}_0^*,\bar{D}_1$),
where the scaled ``$\bar{D}$ mesons'' fields $\bar{D}_v$ and the relativistic fields are related by $\bar{D} =\frac{1}{\sqrt{m}}{\rm e}^{-imv\cdot x}\bar{D}_v $.
In fact, inserting this relation into Lagrangian~(\ref{StartingLagrangian}) and neglecting the subleading terms of $O(1/m)$ corrections, we reproduce the Lagrangian~(\ref{HMET}) successfully.
Eq.~(\ref{StartingLagrangian}) is the basic Lagrangian to be used in the following.

In vacuum, the mean field of $M$ for $\sigma$ meson and pion is $\langle M\rangle_0=f_\pi$ as the spontaneous breaking of chiral symmetry.
Hence, the spin-averaged mass of $(\bar{D},\bar{D}^*)$ and that of $(\bar{D}_0^*,\bar{D}_1)$, respectively, can be parametrized by 
\begin{eqnarray}
M_{(\bar{D},\bar{D}^*)} &=& m-\frac{\Delta_m}{2} \ , \nonumber\\
M_{(\bar{D}_0^*,\bar{D}_1)} &=& m+\frac{\Delta_m}{2}\ ,
\end{eqnarray}
where $\Delta_m$ being the quantity proportional to $f_\pi$, whose form will be given explicitly soon.
Hence, it is essentially important to analyze the value of $\Delta_{m}$ at finite density to investigate the chiral symmetry breaking in nuclear matter.
In vacuum, the value of $\Delta_m$ is given by experimental values as the mass difference between chiral partners: $\Delta_m=M_{(\bar{D}_0^*,\bar{D}_1)}-M_{(\bar{D},\bar{D}^*)}$.
By using the observed ``$\bar{D}$ mesons'' masses $M_{X}$ ($X=\bar{D}$, $\bar{D}^{\ast}$, $\bar{D}_{0}^{\ast}$ and $\bar{D}_{1}$),
we estimate the spin-averaged mass of ($\bar{D},\bar{D}^*$) and that of ($\bar{D}_0^*,\bar{D}_1$) by
\begin{eqnarray}
M_{(\bar{D},\bar{D}^*)} &=& \frac{M_{\bar{D}}+3M_{\bar{D^*}}}{4} \ , \nonumber\\
M_{(\bar{D}_0^*,\bar{D}_1)} &=&\frac{M_{\bar{D}_0^*}+3M_{\bar{D}_1}}{4}\ .
\end{eqnarray}
From the above parametrization, the numerical values of $m$ and $\Delta_m$
are given as
\begin{eqnarray}
m &=& 2190 
\ {\rm MeV} \ , \nonumber\\
\Delta_m &=& 430\ {\rm MeV}\ .
\end{eqnarray}
The parameter $g_A$ in Eq.~(\ref{StartingLagrangian}) is estimated by the decay width of a $D^{\ast}$ meson, $\Gamma[D^*\to D\pi]$, which reads $g_A=0.50$.

In the real world, heavy quark spin symmetry is partly 
 violated by the 
  mass difference between a $\bar{D}$ meson and a $\bar{D}^*$ meson, and by that between a $\bar{D}_0^*$ meson and a $\bar{D}_1$ meson. 
We take into account this violation by shifting the mass of ``$\bar{D}$ mesons'' as 
\begin{eqnarray}
m_{\bar{D}} &=& m-\frac{G_\pi f_\pi}{2} - \frac{\Delta_{\bar{D}}}{2} \ , \nonumber\\
m_{\bar{D}^*} &=& m-\frac{G_\pi f_\pi}{2} + \frac{\Delta_{\bar{D}^*}}{2} \ , \nonumber\\
m_{\bar{D}_0^*} &=& m+\frac{G_\pi f_\pi}{2} - \frac{\Delta_{\bar{D}_0^*}}{2} \ , \nonumber\\
m_{\bar{D}_1} &=& m+\frac{G_\pi f_\pi}{2} + \frac{\Delta_{\bar{D}_1}}{2} \ . \label{MassInVacuum}
\end{eqnarray}
Notice that $\Delta_{m}$ is set to be proportional to $f_{\pi}$, as anticipated, where
$G_\pi$ is defined by $G_\pi=\Delta_m/f_\pi$ whose value is given as $G_\pi = 4.65$. 
We introduce $\Delta_{\bar{D}}$, $\Delta_{\bar{D}^*}$, $\Delta_{\bar{D}_0^*}$ and $\Delta_{\bar{D}_1}$, the mass corrections at ${\cal O}(1/m_{Q})$ from the averaged values, to reproduce the masses of ``$\bar{D}$ mesons'' in vacuum: $m_{\bar{D}}=1869$ MeV, $m_{\bar{D}^*}=2010$ MeV, $m_{\bar{D}_0^*}=2318$ MeV and $m_{\bar{D}_1}=2427$ MeV which lead to $\Delta_{\bar{D}}=202$ MeV, $\Delta_{\bar{D}^*}=80$ MeV, $\Delta_{\bar{D}_0^*}=164$ MeV and $\Delta_{\bar{D}_1}=54$ MeV. 


\section{Modifications of ``$\bar{D}$ mesons'' in nuclear matter}
\label{sec:DInMedium}

In this section, we study the modifications of ``$\bar{D}$ mesons'' in the nuclear matter based on the Lagrangian~(\ref{StartingLagrangian}).  
In particular, we calculate the masses and the spectral functions of $\bar{D}$ $(0^-)$ and $\bar{D}_0^*$ $(0^+)$ mesons at finite density, as the chiral partners, to find some relations between those modifications and partial restoration of chiral symmetry in nuclear matter.
In Sec.~\ref{sec:MassModification}, we investigate the mass modifications of $\bar{D}$ and $\bar{D}_0^*$ mesons in nuclear matter by considering both the mean field and fluctuation of pion and $\sigma$ meson, where the fluctuation is taken by the Hartree-type diagrams (Fig.~\ref{fig:ModifiedMean}).
In Sec.~\ref{sec:spectral_function_D}, including furthermore the fluctuation given by the Fock-type diagrams (Fig.~\ref{fig:DSelfEnergy} and Fig.\ref{fig:D0stSelfEnergy}), we analyze the in-medium spectral functions of the $\bar{D}$ and $\bar{D}_0^*$ mesons.
In Sec.~\ref{sec:Fock_terms}, we summarize the mass shifts of the peaks in the spectral functions.

\subsection{Mass modifications of $\bar{D}$ and $\bar{D}_0^*$ mesons in nuclear matter with Hartree-type diagrams}
\label{sec:MassModification}

\begin{figure*}[htbp]
\centering
\includegraphics*[scale=0.5]{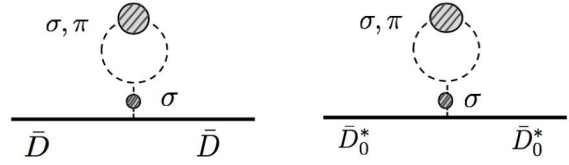}
\caption{Hartree-type one-loop corrections ($\delta\sigma_0^*$) to the self-energies of $\bar{D}$ and $\bar{D}_0^*$ mesons. These corrections provide mass modifications to $\bar{D}$ and $\bar{D}$ mesons, and provide no change in the decay properties. The blobs for a pion and a $\sigma$ meson are defined in Fig.~\ref{fig:ResummedPropagator}.}
\label{fig:ModifiedMean}
\end{figure*}

First of all, we consider the Hartree-type diagram in Fig.~\ref{fig:ModifiedMean}, and investigate the mass shifts of $\bar{D}$ and $\bar{D}_{0}^{\ast}$ mesons.
In this case, there appears no imaginary part in the self-energies, and hence the mass shift includes the only information of the medium modification.
Let us remember that, in vacuum, the mass of ``$\bar{D}$ mesons'' are parametrized by Eqs.~(\ref{MassInVacuum}).
Those relations are changed in nuclear matter due to the interactions between ``$\bar{D}$ mesons'' and nucleons. 
Such modification is supplied by the change of mean field of a sigma meson: $f_\pi\to\sigma_0^*$ in the tree level, where $\sigma_0^*$ is determined by the gap equation (\ref{GapDensDep}) at finite density (cf.~Fig.~\ref{fig:MeanValue}).
 In the mean field level, the masses of ``$\bar{D}$ mesons'' are given as
\begin{eqnarray}
m_{\bar{D}}^* &=& m-\frac{G_\pi \sigma_0^*}{2} - \frac{\Delta_{\bar{D}}}{2} \ , \nonumber\\
m_{\bar{D}^*}^* &=& m-\frac{G_\pi \sigma_0^*}{2} + \frac{\Delta_{\bar{D}^*}}{2} \ , \nonumber\\
m_{\bar{D}_0^*}^* &=& m+\frac{G_\pi \sigma_0^*}{2} - \frac{\Delta_{\bar{D}_0^*}}{2} \ , \nonumber\\
m_{\bar{D}_1}^* &=& m+\frac{G_\pi \sigma_0^*}{2} + \frac{\Delta_{\bar{D}_1}}{2} \ , \label{ModMass}
\end{eqnarray}
by the replacement in Eq.~(\ref{MassInVacuum}).
In the following discussion, we focus on the behaviors of $\bar{D}$ and $\bar{D}_{0}^{\ast}$ mesons, $m_{\bar{D}}^*$ and $m_{\bar{D}_0^*}^*$.

In the present study, in addition to the replacement from $f_{\pi}$ to $\sigma_{0}^{\ast}$ in the tree level, we further compute the one-loop corrections. The diagrams of the one-loop corrections to self-energies of $\bar{D}$ and $\bar{D}_0^*$ mesons are shown in Figs.~\ref{fig:ModifiedMean},~\ref{fig:DSelfEnergy} and~\ref{fig:D0stSelfEnergy}. 
The one-loop diagrams in Fig.~\ref{fig:ModifiedMean} correspond to the Hartree-type corrections which generate the correction term $\delta\sigma_0^*$ to mean field: $\sigma_0^*\to\sigma_0^*+\delta\sigma_0^*$.
This correction provides mass modifications to $\bar{D}$ and $\bar{D}_0^*$ mesons.
The one-loop diagrams in Figs.~\ref{fig:DSelfEnergy} and~\ref{fig:D0stSelfEnergy} correspond to the Fock-type corrections which provide, not only mass modifications in the real part of the self-energies of $\bar{D}$ and $\bar{D}_0^*$ mesons, but also imaginary part leading to the modifications of the decay properties.
It will turn out, in fact, that these Fock-type diagrams are significant in the spectral functions for $\bar{D}$ and $\bar{D}_0^*$ meson channels.

In the Hartree-type diagrams (Fig.~\ref{fig:ModifiedMean}), 
the correction term $\delta\sigma_0^*$ can be calculated as
\begin{widetext}
\begin{eqnarray}
\delta\sigma_0^* &=& -\frac{3\lambda\sigma_0^*}{\tilde{m}_\sigma^{*2}}\int\frac{d^4k}{(2\pi)^4}\left(F(\vec{k};\Lambda)\right)^2\left(\tilde{G}^*_\sigma(k_0,\vec{k})-\tilde{G}^{\rm vac}_\sigma(k_0,\vec{k})\right) \nonumber\\
&&-\frac{3\lambda\sigma_0^*}{\tilde{m}_\sigma^{*2}}\int\frac{d^4k}{(2\pi)^4}\left(F(\vec{k};\Lambda)\right)^2\left(\tilde{G}^*_\pi(k_0,\vec{k})-\tilde{G}_\pi^{\rm vac}(k_0,\vec{k})\right) \nonumber\\ \nonumber\\
&=& \frac{3\lambda\sigma_0^*}{\tilde{m}_\sigma^{*2}}\int\frac{d^4k}{(2\pi)^4}\left(F(\vec{k};\Lambda)\right)^2{\rm Im}\left[\frac{1}{k^2-m_\sigma^{*2}-i\tilde{\Sigma}_\sigma^*(k_0,\vec{k})} -\frac{1}{k^2-m_\sigma^2+i\epsilon}\right]\nonumber\\
&&+\frac{3\lambda\sigma_0^*}{\tilde{m}_\sigma^{*2}}\int\frac{d^4k}{(2\pi)^4}\left(F(\vec{k};\Lambda)\right)^2{\rm Im}\left[\frac{1}{k^2-m_\pi^{*2}-i\tilde{\Sigma}_\pi^*(k_0,\vec{k})}-\frac{1}{k^2-m_\pi^2+i\epsilon}\right] \nonumber\\ \nonumber\\
&=& -\frac{3\lambda\sigma_0^*}{2\tilde{m}_\sigma^{*2}} \int\frac{d^4k}{(2\pi)^4} \left(F(\vec{k};\Lambda)\right)^2\epsilon(k_0)\left\{\rho_{\sigma}^*(k_0,\vec{k})-\rho^{\rm vac}_\sigma(k)\right\} \nonumber\\ 
&& -\frac{3\lambda\sigma_0^*}{2\tilde{m}_\sigma^{*2}} \int\frac{d^4k}{(2\pi)^4}\left(F(\vec{k};\Lambda)\right)^2 \epsilon(k_0)\left\{\rho^*_{\pi}(k_0,\vec{k})-\rho^{\rm vac}_\pi(k_0,\vec{k})\right\}\ .\label{LocalOneLoop} \nonumber\\
\end{eqnarray}
\end{widetext}
In the first line in Eq.~(\ref{LocalOneLoop}), $\tilde{G}_{\pi(\sigma)}^*(q_0,\vec{q})$ is the in-medium propagator defined by $\tilde{G}_{\pi(\sigma)}^*(q_0,\vec{q})=i\left(\tilde{\Gamma}^{(2)*}_{\pi(\sigma)}(q_0,\vec{q})\right)^{-1}$, and $\tilde{G}_{\pi(\sigma)}^{\rm vac}(q_0,\vec{q})$ is the propagator in the vacuum. In obtaining Eq.~(\ref{LocalOneLoop}), we have used the property that $\delta\sigma_0^*$ is real, and Eqs.~(\ref{RelationRe}) and~(\ref{RelationIm}). The spectral functions for $\rho_\sigma^*(q_0,\vec{q})$ and $\rho_\pi^{*}(q_0,\vec{q})$ have already obtained in Eq.~(\ref{RenormalizedSpectral}).
$\tilde{m} _{\sigma}^{\ast}$ is the on-shell mass of $\sigma$ meson in nuclear matter defined by $\tilde{\Gamma}^{*(2)}_\sigma(\tilde{m}_\sigma^{\ast},\vec{0})=0$. $F(\vec{k};\Lambda)$ is the form factor in momentum space parametrized by
\begin{eqnarray}
F(\vec{k};\Lambda) = \frac{\Lambda^2}{|\vec{k}|^2+\Lambda^2} \ ,
\end{eqnarray}
with the cutoff parameter $\Lambda$,
which is introduced 
for the finite size effect of hadrons.
The value of $\Lambda$ is a free value, and 
we set $\Lambda=300$ MeV as a typical low energy scale, which is comparable to, but slightly larger than the Fermi momentum at which we shall study. 
In Eq.~(\ref{LocalOneLoop}), we subtracted the contributions in vacuum: $\tilde{G}_{\pi(\sigma)}^{\rm vac}$ or $\rho^{\rm vac}_{\pi(\sigma)}(k_0,\vec{k}) = 2\pi \epsilon(k_0)\delta(k^2-m_{\pi(\sigma)}^{2})$.
This is a necessary procedure for the condition that $\delta \sigma_{0}^{\ast}$ vanishes 
 in vacuum.
 In this respect, we regard $\sigma_0^*$ as the mean field renormalized in vacuum.
 
\begin{figure}[htbp]
\centering
\includegraphics*[scale=0.7]{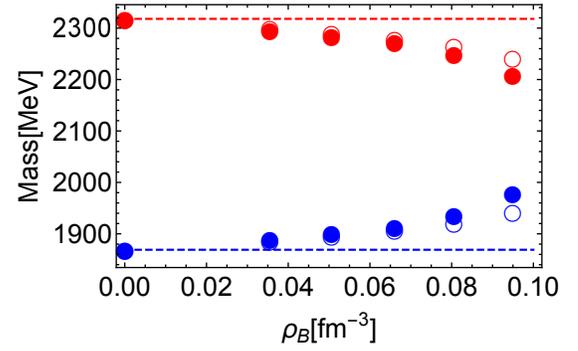}
\caption{(color online) Density dependence of masses of $\bar{D}$ and $\bar{D}_0^*$ mesons with one-loop corrections in Fig.~\ref{fig:ModifiedMean}. Blue (red) circles show the mass of $\bar{D}$ ($\bar{D}_0^*$) meson at each density. Filled circles are the results with mean field and one-loop corrections $\sigma_0^*+\delta\sigma_0^*$. Open circles show the masses of $\bar{D}$ and $\bar{D}_0^*$ mesons with only mean field $\sigma_0^*$ plotted as a reference. The dashed lines refer the mass of $\bar{D}$ and $\bar{D}_0^*$ mesons in the vacuum. }
\label{fig:MassOfDWithModifiedMean}
\end{figure}

 With the above setup, the mass modifications to $\bar{D}$ and $\bar{D}_0^*$ mesons by one-loop diagrams in Fig.~\ref{fig:ModifiedMean} are obtained as
\begin{eqnarray}
\hat{m}^*_{\bar{D}} &=& m-\frac{G_\pi(\sigma_0^*+\delta\sigma_0^*)}{2}-\frac{\Delta_{\bar{D}}}{2} \ , \nonumber\\
\hat{m}^*_{\bar{D}_0^*} &=& m+\frac{G_\pi(\sigma_0^*+\delta\sigma_0^*)}{2}-\frac{\Delta_{\bar{D}_0^*}}{2}\ .
\end{eqnarray}
We present the density dependence of the masses of the $\bar{D}$ and $\bar{D}_0^*$ mesons in Fig.~\ref{fig:MassOfDWithModifiedMean}.
\begin{figure}[htbp]
\centering
\includegraphics*[scale=0.7]{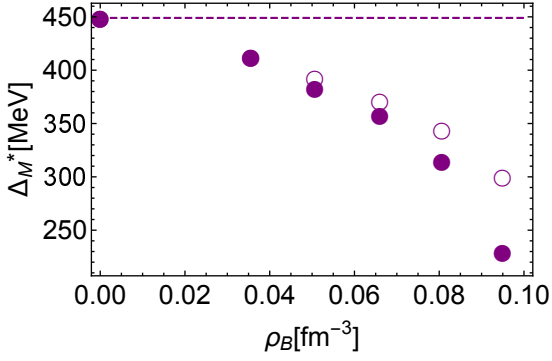}
\caption{(color online) Density dependence of mass difference between $\bar{D}_0^*$ and $\bar{D}$ mesons with one-loop corrections, $\Delta_{m}^{\ast}=m_{\bar{D}_{0}^{\ast}}^{\ast}-m_{\bar{D}}^{\ast}$
 for $m_{\bar{D}_{0}^{\ast}}^{\ast}$ and $m_{\bar{D}}^{\ast}$ being shown in Fig.~\ref{fig:ModifiedMean}.
Filled purple circles indicate the mass difference when both mean field and one-loop correction, $\sigma_0^*+\delta\sigma_0^*$, are considered, and open circles indicate the mass difference when only the mean field $\sigma_0^*$ is considered. The dashed line refers the mass difference between $\bar{D}_0^*$ and $\bar{D}$ mesons in the vacuum.}
\label{fig:DeltaMWithModifiedMean}
\end{figure}
Blue (red) circles indicate the mass of $\bar{D}$ ($\bar{D}_0^*$) meson.  
Filled circles indicate the results with the mean field and the one-loop corrections by $\sigma_0^*+\delta\sigma_0^*$.
Open circles show the masses of $\bar{D}$ and $\bar{D}_0^*$ mesons with only mean field $\sigma_0^*$ 
 as a reference to see the effect of $\delta\sigma_0^*$. 
The increase (decrease) of the masses of the $\bar{D}$ ($\bar{D}_0^*$) meson indicate the partial restoration of chiral symmetry at finite density, because the mass splitting between a $\bar{D}$ meson and a $\bar{D}_0^*$ should be proportional to the expectation value of the $\sigma$ meson field, as stated already.
Interestingly, we find that the mass of $\bar{D}$ ($\bar{D}_0^*$) mesons with one-loop corrections are increased (decreased) more rapidly than the case when we consider only the mean field $\sigma_0^*$.
 This result indicates that the one-loop correction in Fig.~\ref{fig:ModifiedMean} accelerates the restoration of chiral symmetry in the nuclear matter. 
The mass difference between $\bar{D}$ and $\bar{D}_0^*$ meson at finite density, defined by $\Delta_m^*=m_{\bar{D}_{0}^{\ast}}^{\ast}-m_{\bar{D}}^{\ast}$, is shown in Fig.~\ref{fig:DeltaMWithModifiedMean}.
 Filled purple circles indicate the mass difference with mean field and one-loop correction $\sigma_0^*+\delta\sigma_0^*$, and open circles indicate the mass difference with mean field $\sigma_0^*$ only.
We see again that the $\Delta_{m}^{\ast}$ in the former case becomes smaller than that in the latter case.

\subsection{Spectral function for $\bar{D}_0^*$ and $\bar{D}$ meson channel}
\label{sec:spectral_function_D}
\begin{figure*}[htbp]
\centering
\includegraphics*[scale=0.8]{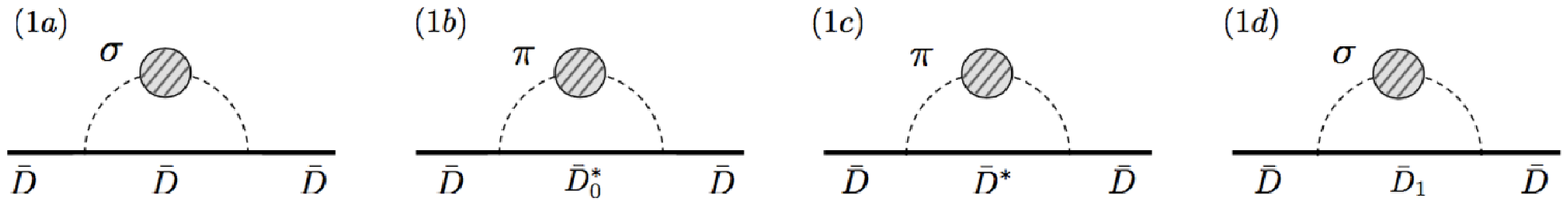}
\caption{Fock-type one-loop corrections to the self-energy for $\bar{D}$ meson ($\tilde{\Sigma}_{\bar{D}}^{*}(q_0,\vec{q})$).}
\label{fig:DSelfEnergy}
\end{figure*}

\begin{figure*}[htbp]
\centering
\includegraphics*[scale=0.8]{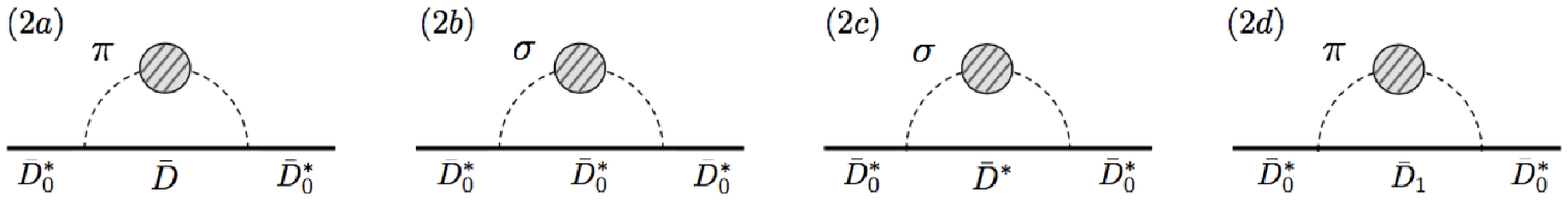}
\caption{Fock-type one-loop corrections to the self-energy for $\bar{D}_0^\ast$ meson ($\tilde{\Sigma}_{\bar{D}_0^*}^{*}(q_0,\vec{q})$).}
\label{fig:D0stSelfEnergy}
\end{figure*}

Next, we take into account the one-loop corrections to $\bar{D}$ and $\bar{D}_0^*$ by considering not only the diargams in Fig.~\ref{fig:ModifiedMean} but also ones in Fig.~\ref{fig:DSelfEnergy} and Fig.~\ref{fig:D0stSelfEnergy}, and investigate the spectral function for each $\bar{D}$ 
channel and $\bar{D}_0^*$ 
 channel. 


We calculate the self-energies in Fig.~\ref{fig:DSelfEnergy} and~\ref{fig:D0stSelfEnergy} in the following procedure.
First, we calculate the imaginary part of the retarded self-energy ${\rm Im}\tilde{\Sigma}_{\bar{D}}^{*R}(q_0,\vec{q})$ and ${\rm Im}\tilde{\Sigma}_{\bar{D}_0^*}^{*R}(q_0,\vec{q})$. 
Their imaginary part is not suffered by any ultraviolet (UV) divergences, and hence there is no need to introduce the momentum cutoff.
Secondly, to obtain the real part of retarded self-energy ${\rm Re}\tilde{\Sigma}_{\bar{D}}^{*R}(q_0,\vec{q})$ and ${\rm Re}\tilde{\Sigma}_{\bar{D}_0^*}^{*R}(q_0,\vec{q})$, we employ the subtracted dispersion relation
\begin{eqnarray}
{\rm Re}\tilde{\Sigma}_{\bar{D}}^{*R}(q_0,\vec{q}) &=& \frac{q^2-m_{\bar{D}}^2}{\pi} \nonumber\\
&& \times {\rm P} \!\! \int_{0}^\infty \!\!\!\! dz^2\frac{{\rm Im}\tilde{\Sigma}_{\bar{D}}^{*R}(z,\vec{q})}{(z^2-q_0^2)(z^2-|\vec{q}|^2-m_{\bar{D}}^2)} \ , \nonumber\\ 
{\rm Re}\tilde{\Sigma}_{\bar{D}_0^*}^{*R}(q_0,\vec{q}) &=& \frac{q^2-m_{\bar{D}_0^*}^2}{\pi} \nonumber\\
&&\times {\rm P} \!\! \int_{0}^\infty \!\!\!\! dz^2\frac{{\rm Im}\tilde{\Sigma}_{\bar{D}_0^*}^{*R}(z,\vec{q})}{(z^2-q_0^2)(z^2-|\vec{q}|^2-m_{\bar{D}_0^*}^2)} \ ,  \nonumber\\  \label{RealPartD}
\end{eqnarray}
where the UV divergences are regularized automatically.
Here, $z$ is the complex variable on the complex-energy plane, and ${\rm P}$ stands for the principal value integral.
The subtracted dispersion relation in Eq.~(\ref{RealPartD}) are derived in Appendix.~\ref{sec:SubDispersion}.
Note that 
 $m_{\bar{D}}$ and $m_{\bar{D}_0^*}$ are the renormalized masses of $\bar{D}$ and $\bar{D}_0^*$ meson in vacuum. 
 
One of the most useful way to compute the imaginary part of retarded self-energy is the ``spectral representation'' method~\cite{TFT}. Here, we shall show the detailed calculation of the imaginary part of retarded self-energy in Fig.~\ref{fig:DSelfEnergy} (1a) as an example. According to Eq.~(\ref{ImRelationSigma}), the retarded self-energy ${\rm Im}\tilde{\Sigma}^{*R}_{\bar{D}(1a)}(q_0,\vec{q})$ satisfies following relation
\begin{eqnarray}
{\rm Im}\tilde{\Sigma}^{*R}_{\bar{D}(1a)}(q_0,\vec{q}) = \frac{1}{2}\left(\tilde{\Sigma}_{\bar{D}(1a)}^{*>}(q_0,\vec{q})-\tilde{\Sigma}_{\bar{D}(1a)}^{*<}(q_0,\vec{q})\right)  \ .\nonumber\\ \label{ImRel}
\end{eqnarray}
$\tilde{\Sigma}^{*>}_{\bar{D}(1a)}(q_0,\vec{q})$ and $\tilde{\Sigma}^{*<}_{\bar{D}(1a)}(q_0,\vec{q})$ are the Fourier transformation of the greater self-energy $\Sigma^{*>}_{\bar{D}(1a)}(x_0,\vec{x})$ and lesser self-energy $\Sigma_{\bar{D}(1a)}^{*<}(x_0,\vec{x})$, respectively, in the coordinate space. The $\Sigma^{*>}_{\bar{D}(1a)}(x_0,\vec{x})$ and $\Sigma_{\bar{D}(1a)}^{*<}(x_0,\vec{x})$ are defined through the self-energy $\Sigma_{\bar{D}(1a)}^{*}(x_0,\vec{x})$ by
\begin{eqnarray}
&&\Sigma_{\bar{D}(1a)}^*(x_0,\vec{x}) \nonumber\\
&&\ \ \ \ \ \ \ = \theta(x_0)\Sigma^{*>}_{\bar{D}(1a)}(x_0,\vec{x}) + \theta(-x_0)\Sigma^{*<}_{\bar{D}(1a)}(x_0,\vec{x})\ , \nonumber\\ \label{SelfEDeco}
\end{eqnarray}
where $\theta(x_0)$ is the Heaviside's step function. $\tilde{\Sigma}^{*R}_{\bar{D}(1a)}(q_0,\vec{q})$ is the Fourier transformation of the retarded self-energy ${\Sigma}^{*R}_{\bar{D}(1a)}(x_0,\vec{x})$ which are defined by
\begin{eqnarray}
{\Sigma}^{*R}_{\bar{D}(1a)}(x_0,\vec{x}) \equiv i\theta(x_0)\left(\Sigma^{*>}_{\bar{D}(1a)}(x_0,\vec{x})-\Sigma^{*<}_{\bar{D}(1a)}(x_0,\vec{x})\right)\ . \nonumber\\
\end{eqnarray}

In order to calculate ${\rm Im}\tilde{\Sigma}^{*R}_{\bar{D}(1a)}(q_0,\vec{q})$, as shown in Eq.~(\ref{ImRel}), we need to get $\tilde{\Sigma}_{\bar{D}(1a)}^{*>}(q_0,\vec{q})$ and $\tilde{\Sigma}_{\bar{D}(1a)}^{*<}(q_0,\vec{q})$, and these are obtained through the self-energy $\tilde{\Sigma}_{\bar{D}(1a)}^*(q_0,\vec{q})$ as in Eq.~(\ref{SelfEDeco}). To start, we should evaluate $\Sigma^*_{\bar{D} (1a)}(x_0,\vec{x})$.  $\Sigma^*_{\bar{D} (1a)}(x_0,\vec{x})$ is calculated from the Lagrangian~(\ref{StartingLagrangian}) as
\begin{widetext}
\begin{eqnarray}
\Sigma^*_{\bar{D} (1a)}(x_0,\vec{x}) &=&  \left(imG_\pi\right)^2G_{\bar{D}}(x_0,\vec{x})G_\sigma^*(x_0,\vec{x}) \nonumber\\
&=& \left(imG_\pi\right)^2\left[\theta(x_0)G^>_{\bar{D}}(x_0,\vec{x})+\theta(-x_0)G_{\bar{D}}^<(x_0,\vec{x})\right]\left[\theta(x_0)G^{*>}_{\sigma}(x_0,\vec{x})+\theta(-x_0)G_{\sigma}^{*<}(x_0,\vec{x})\right]\nonumber\\
&=& \left(imG_\pi\right)^2\left[\theta(x_0)G_{\bar{D}}^>(x_0,\vec{x})G_\sigma^{*>}(x_0,\vec{x})+\theta(-x_0)G_{\bar{D}}^<(x_0,\vec{x})G_\sigma^{*<}(x_0,\vec{x})\right] \label{SigmaAB}\ , 
\end{eqnarray}
where we have defined the greater Green's functions $G^>_{\bar{D}}(x_0,\vec{x})$, $G^
{*>}_{\sigma}(x_0,\vec{x})$ and the lesser Green's functions $G^<_{\bar{D}}(x_0,\vec{x})$, $G^{*<}_{\sigma}(x_0,\vec{x})$ through the two-point functions of $\bar{D}$ meson and $\sigma$ meson by
\begin{eqnarray}
G_{\sigma}^{*}(x_0,\vec{x}) &=& \theta(x_0)G_{\sigma}^{*>}(x_0,\vec{x}) + \theta(-x_0)G_{\sigma}^{*<}(x_0,\vec{x}) \nonumber\\
G_{\bar{D}}(x_0,\vec{x}) &=& \theta(x_0)G_{\bar{D}}^>(x_0,\vec{x}) + \theta(-x_0)G_{\bar{D}}^{<}(x_0,\vec{x}) \ .
\end{eqnarray}
Comparing Eq.~(\ref{SigmaAB}) with Eq.~(\ref{SelfEDeco}), we get
\begin{eqnarray}
\Sigma_{\bar{D}(1a)}^>(x_0,\vec{x}) &=&  \left(imG_\pi\right)^2G_{\bar{D}}^>(x_0,\vec{x})G_\sigma^{*>}(x_0,\vec{x}) \nonumber\\
\Sigma_{\bar{D}(1a)}^<(x_0,\vec{x}) &=& \left(imG_\pi\right)^2G_{\bar{D}}^<(x_0,\vec{x})G_\sigma^{*<}(x_0,\vec{x}) \ ,
\end{eqnarray}
and Fourier transformation reads
\begin{eqnarray}
\tilde{\Sigma}_{\bar{D}(1a)}^>(q_0,\vec{q}) &=& \left(imG_\pi\right)^2\int\frac{d^4k}{(2\pi)^4}\left(F(\vec{k};\Lambda)\right)^2\tilde{G}_{\bar{D}}^>(k_0,\vec{k})\tilde{G}_\sigma^{*>}(q_0-k_0,\vec{q}-\vec{k}) \nonumber\\
\tilde{\Sigma}_{\bar{D}(1a)}^<(q_0,\vec{q}) &=& \left(imG_\pi\right)^2\int\frac{d^4k}{(2\pi)^4}\left(F(\vec{k};\Lambda)\right)^2\tilde{G}_{\bar{D}}^<(k_0,\vec{k})\tilde{G}_\sigma^{*<}(q_0-k_0,\vec{q}-\vec{k}) \ , \label{FTSelfE}
\end{eqnarray}
where $\tilde{G}^>_{\bar{D}}(q_0,\vec{q})$, $\tilde{G}_\sigma^{*>}(q_0,\vec{q})$ and $\tilde{G}_{\bar{D}}^<(q_0,\vec{q})$, $\tilde{G}^{*<}_{\sigma}(q_0,\vec{q})$ are the Fourier transformation of $G^>_{\bar{D}}(x_0,\vec{x})$, $G^
{*>}_{\sigma}(x_0,\vec{x})$ and $G^<_{\bar{D}}(x_0,\vec{x})$, $G^{*<}_{\sigma}(x_0,\vec{x})$, respectively. In obtaining Eq.~(\ref{FTSelfE}), we have inserted the form factor $F(\vec{k};\Lambda)$ for each vertexes. According to Eq.~(\ref{GLRel}), $\tilde{G}^>_{\bar{D}}(q_0,\vec{q})$, $\tilde{G}^{*>}_{\sigma}(q_0,\vec{q})$, $\tilde{G}^<_{\bar{D}}(q_0,\vec{q})$ and $\tilde{G}^{*<}_{\sigma}(q_0,\vec{q})$ take
\begin{eqnarray}
\tilde{G}^>_{\bar{D}}(q_0,\vec{q}) = \theta(q_0)\rho_{\bar{D}}(q_0,\vec{q}) \ \ &,&\ \ \  \tilde{G}^{*>}_{\sigma}(q_0,\vec{q}) = \theta(q_0)\rho_{\sigma}^*(q_0,\vec{q})\ , \nonumber\\
\tilde{G}^<_{\bar{D}}(q_0,\vec{q}) = -\theta(-q_0)\rho_{\bar{D}}(q_0,\vec{q}) \ \ &,&\ \ \ 
\tilde{G}^{*<}_{\sigma}(q_0,\vec{q}) = -\theta(q_0)\rho^*_{\sigma}(q_0,\vec{q}) \ . \label{GLForm}
\end{eqnarray}
$\rho_\sigma^*(q_0,\vec{q})$ is the spectral function for $\sigma$ meson which has been obtained in Eq.~(\ref{RenormalizedSpectral}). $\rho_{\bar{D}}(k_0,\vec{k})$ is the spectral function for $\bar{D}$ meson
\begin{eqnarray}
\rho_{\bar{D}}(k_0,\vec{k}) = 2\pi\epsilon(k_0)\delta(k^2-m_{\bar{D}}^{*2})\ ,
\end{eqnarray}
which takes the form of a free particle, but its mass is modified by mean field $\sigma_0^*$ in Eq.~(\ref{ModMass}). Namely, the mass of $\bar{D}$ meson in the internal line in Fig.~\ref{fig:DSelfEnergy} $(1a)$ is modified by mean field $\sigma_0^*$, and perturbation series are defined around this mean field level. Then, by using Eqs.~(\ref{ImRel}) and~(\ref{GLForm}), ${\rm Im}\tilde{\Sigma}^{*R}_{\bar{D}(1a)}(q_0,\vec{q})$ is finally calculated as
\begin{eqnarray}
{\rm Im}\tilde{\Sigma}_{\bar{D}(1a)}^R(q_0,\vec{q}) &=& \frac{1}{2}\left(\tilde{\Sigma}_{\bar{D}(1a)}(q_0,\vec{q})-\tilde{\Sigma}_{\bar{D}(1a)}^<(q_0,\vec{q})\right) \nonumber\\
&=& \frac{1}{2}\left(imG_\pi\right)^2\int\frac{d^4k}{(2\pi)^4}\left(F(\vec{k};\Lambda)\right)^2\tilde{G}_{\bar{D}}^>(k_0,\vec{k})\tilde{G}_\sigma^{*>}(q_0-k_0,\vec{q}-\vec{k}) \nonumber\\
&& -  \frac{1}{2}\left(imG_\pi\right)^2\int\frac{d^4k}{(2\pi)^4}\left(F(\vec{k};\Lambda)\right)^2\tilde{G}_{\bar{D}}^<(k_0,\vec{k})\tilde{G}_\sigma^{*<}(q_0-k_0,\vec{q}-\vec{k})  \nonumber\\
&=& \frac{1}{2}\left(imG_\pi\right)^2\int\frac{d^4k}{(2\pi)^4}\left(F(\vec{k};\Lambda)\right)^2\theta(k_0)\rho_{\bar{D}}(k_0,\vec{q})\theta(q_0-k_0)\rho_{\sigma}^*(q_0-k_0,\vec{q}-\vec{k})\nonumber\\
&& -  \frac{1}{2}\left(imG_\pi\right)^2\int\frac{d^4k}{(2\pi)^4}\left(F(\vec{k};\Lambda)\right)^2\theta(-k_0)\rho_{\bar{D}}(k_0,\vec{k})\theta(-q_0+k_0)\rho_\sigma^*(q_0-k_0,\vec{q}-\vec{k})\nonumber\\
&=&-\frac{1}{2}m^2G_\pi^2\int\frac{d^3k}{(2\pi)^3}\left(F(\vec{k};\Lambda)\right)^2 \nonumber\\
&& \times \frac{1}{2E_k^{\bar{D}}}\left\{\theta(q_0-E_k^{\bar{D}})\rho^*_\sigma(q_0-E_k^{\bar{D}},\vec{q}-\vec{k})+\theta(-E_k^{\bar{D}}-q_0)\rho^*_\sigma(q_0+E_k^{\bar{D}},\vec{q}-\vec{k})\right\}  \ . \label{ImSigmaComp}
\end{eqnarray}
At the last line in Eq.~(\ref{ImSigmaComp}), we have defined $E_k^{\bar{D}}\equiv\sqrt{|\vec{k}|^2+m_{\bar{D}}^{*2}}$.

The imaginary part of the retarded self-energy of other diagrams in Figs.~\ref{fig:DSelfEnergy} and~\ref{fig:D0stSelfEnergy}: ${\rm Im}\tilde{\Sigma}_{\bar{D}(1a)}^{*R}(q_0,\vec{q})$, ${\rm Im}\tilde{\Sigma}_{\bar{D}(1b)}^{*R}(q_0,\vec{q})$, ${\rm Im}\tilde{\Sigma}_{\bar{D}(1c)}^{*R}(q_0,\vec{q})$, ${\rm Im}\tilde{\Sigma}_{\bar{D}(1d)}^{*R}(q_0,\vec{q})$ and ${\rm Im}\tilde{\Sigma}_{\bar{D}_0^*(2a)}^{*R}(q_0,\vec{q})$, ${\rm Im}\tilde{\Sigma}_{\bar{D}_0^*(2b)}^{*R}(q_0,\vec{q})$, ${\rm Im}\tilde{\Sigma}_{\bar{D}_0^*(2c)}^{*R}(q_0,\vec{q})$, ${\rm Im}\tilde{\Sigma}_{\bar{D}_0^*(2d)}^{*R}(q_0,\vec{q})$, can be calculated in the same way, and they are summarized in Appendix.~\ref{sec:ImaginaryParts}.  Then defining 
\begin{eqnarray}
{\rm Im}\Sigma_{\bar{D}}^{*R}(q_0,\vec{q}) &\equiv& {\rm Im}\Sigma_{\bar{D}(1a)}^{*R}(q_0,\vec{q})+{\rm Im}\Sigma_{\bar{D}(1b)}^{*R}(q_0,\vec{q})+{\rm Im}\Sigma_{\bar{D}(1c)}^{*R}(q_0,\vec{q})+{\rm Im}\Sigma_{\bar{D}(1d)}^{*R}(q_0,\vec{q}) \ , \\
{\rm Im}\Sigma_{\bar{D}_0^*}^{*R}(q_0,\vec{q}) &\equiv&  {\rm Im}\Sigma_{\bar{D}_0^*(2a)}^{*R}(q_0,\vec{q})+{\rm Im}\Sigma_{\bar{D}_0^*(2b)}^{*R}(q_0,\vec{q})+{\rm Im}\Sigma_{\bar{D}_0^*(2c)}^{*R}(q_0,\vec{q})+{\rm Im}\Sigma_{\bar{D}_0^*(2d)}^{*R}(q_0,\vec{q})\ , 
\end{eqnarray}
and utilizing the subtracted dispersion relation~(\ref{RealPartD}), we can get the spectral functions for $\bar{D}$ and $\bar{D}_0^*$ mesons 
\begin{eqnarray}
\rho_{\bar{D}}^*(q_0,\vec{q}) &=& \frac{-2{\rm Im}\Sigma_{\bar{D}}^{*R}(q_0,\vec{q})}{\left[q^2-\hat{m}_{\bar{D}}^2-{\rm Re}\Sigma_{\bar{D}}^{*R}(q_0,\vec{q})\right]^2+\left[{\rm Im}\Sigma_{\bar{D}}^{*R}(q_0,\vec{q})\right]^2} \ ,\\ \nonumber\\
\rho_{\bar{D}_0^*}^*(q_0,\vec{q}) &=& \frac{-2{\rm Im}\Sigma_{\bar{D}_0^*}^{*R}(q_0,\vec{q})}{\left[q^2-\hat{m}_{\bar{D}_0^*}^2-{\rm Re}\Sigma_{\bar{D}_0^*}^{*R}(q_0,\vec{q})\right]^2+\left[{\rm Im}\Sigma_{\bar{D}_0^*}^{*R}(q_0,\vec{q})\right]^2} \ .
\end{eqnarray}
\end{widetext}

\begin{figure*}[htbp]
\centering
\includegraphics*[scale=0.62]{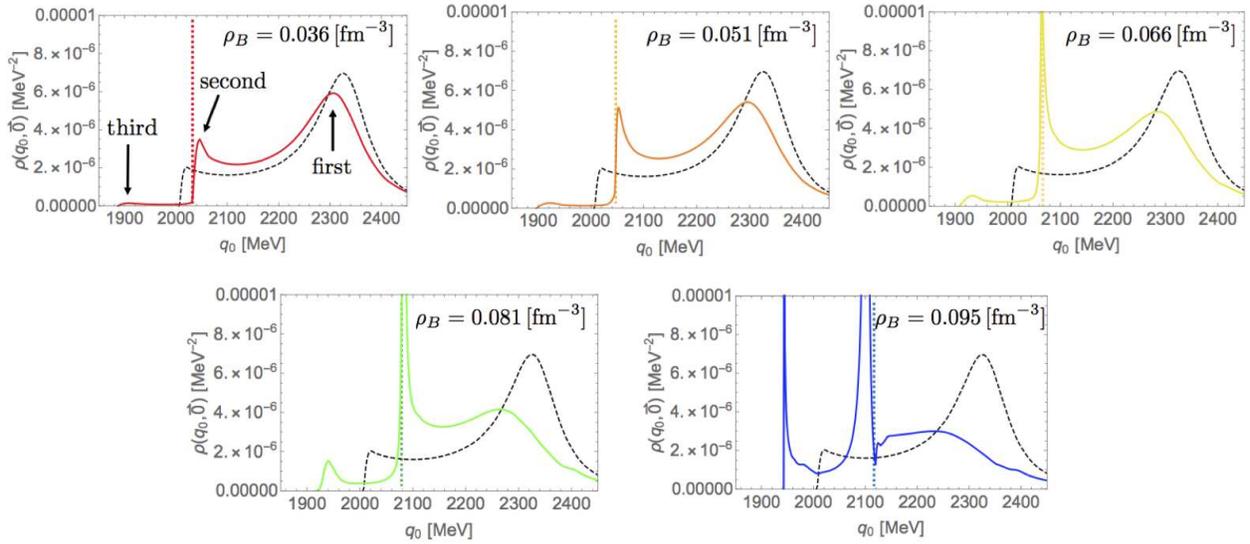}
\caption{(color online) Spectral function for $\bar{D}_0^*$ meson channel at several densities. Dashed black curve indicates the spectral function in vacuum, and vertical dotted line is the threshold of $\bar{D}+\pi$. We can find three peaks as shown in the figure at $\rho_B=0.036$ [fm$^{-3}$]. The detailed explanation of these peaks are given in the text.}
\label{fig:SpectralEachKf}
\end{figure*}
\begin{figure*}[htbp]
\centering
\includegraphics*[scale=0.63]{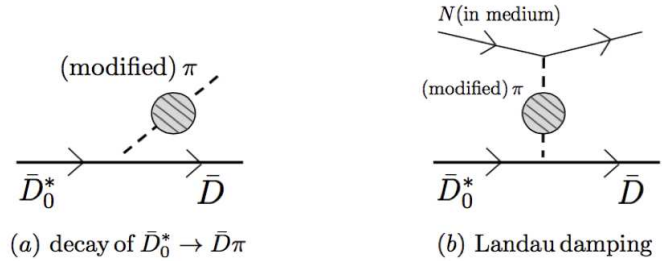}
\caption{Diagrammatical interpretation for $(a)$ decay process of $\bar{D}_0^*\to\bar{D}\pi$ and $(b)$ Landau damping.}
\label{fig:DDiagrams}
\end{figure*}

First of all, we plot the spectral function for $\bar{D}_0^*$ channel with $\vec{q}=\vec{0}$ at several baryon number densities in Fig.~\ref{fig:SpectralEachKf}. 
Colored curves show the obtained results, and dashed black curve is the spectral function in the vacuum. Vertical dotted line is the threshold of $\bar{D}+\pi$ at mean field level. From this figure, we can find three peaks and density dependence of them differs each others. The first peak from right essentially corresponds to the imaginary part induced by the decay process into $\bar{D}$ meson by emitting a pion as shown in Fig.~\ref{fig:DDiagrams} $(a)$, i.e., this corresponds to the resonance of $\bar{D}_0^*$ meson state. As density increases, the peak position gradually shifts to lower energy. This change is caused since the masses of $\bar{D}_0^*$ and $\bar{D}$ meson get close as density increases, which is a consequence of partial restoration of chiral symmetry. Besides, the height of this peak gets small, and especially at $\rho_B=0.095$ [fm$^{-3}$] (blue colored plot), we cannot see the bump structure well. This is originated from the enlargement of imaginary part due to the collisional broadening. Second peak from right is induced by the threshold enhancement effect. This peak grows remarkably as density increases and the peak position shifts to higher energy contrasts to the first peak from right. A similar tendency in spectral function for $\sigma$ meson at temperature was found in Ref.~\cite{Chiku:1997va}. In this reference, the spectral function for $\sigma$ meson at finite temperature was studied within a framework of $O(4)$ linear $\sigma$ model, and a threshold enhancement was shown. This result was regarded as a noticeable precritical phenomenon of the chiral phase transition. Similarly, the second peak from right in our result is the strong consequence of the partial restoration of chiral symmetry in nuclear matter. Third peak from right can be interpreted as the Landau damping. Landau damping is seen in the space-like domain in spectral function for $\sigma$ meson and pion as shown in Fig.~\ref{fig:Spectrals}, and is diagrammatically understood as the scattering process in Fig.~\ref{fig:LandauBump}. In case of $\bar{D}_0^*$ meson, this effect is drawn diagrammatically in Fig.~\ref{fig:DDiagrams} $(b)$. This is the nuclear matter effect and its peak grows as the density increases. 

\begin{figure}[htbp]
\centering
\includegraphics*[scale=0.7]{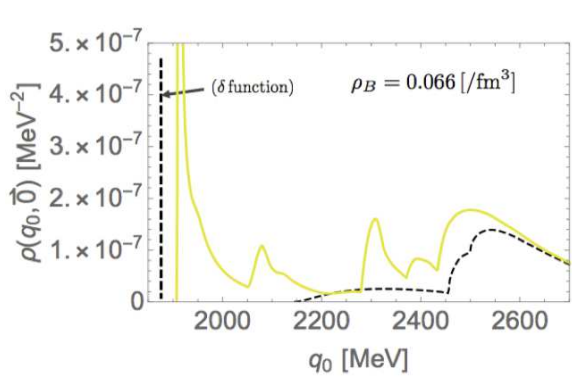}
\caption{(color online) Spectral function for $\bar{D}$ meson channel at $\rho_B=0.066$ [fm$^{-3}$]. Dashed black curve is the spectral function for $\bar{D}$ meson channel in the vacuum.}
\label{fig:SpectralD}
\end{figure}
In Fig.~\ref{fig:SpectralD}, we plot the spectral function for $\bar{D}$ meson channel at $\rho_B=0.066$ [fm$^{-3}$] by yellow curve, and that in the vacuum by dashed black curve. Although the delta function which indicates one particle state for $\bar{D}$ meson pole stands at $q_0=1869$ [MeV] in the vacuum, this peak is smeared at $\rho_B=0.066$ [fm$^{-3}$] due to the Landau damping as shown the first peak from left in Fig.~\ref{fig:SpectralD}. Some various bump structure is observed in the spectral function for $\bar{D}$ meson channel, however, their magnitude is smaller than the one in the spectral function for $\bar{D}_0^*$ meson channel.

\subsection{Summary of the peak shift of $\bar{D}$ and $\bar{D}_0^*$ mesons with Fock-type corrections as well as Hartree-type corrections}
\label{sec:Fock_terms}

\begin{figure}[htbp]
\centering
\includegraphics*[scale=0.7]{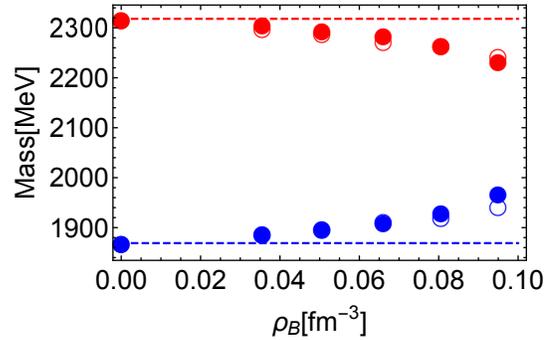}
\caption{(color online) Density dependence of mass of $\bar{D}$ and $\bar{D}_0^*$ mesons with all one-loop corrections in Figs.~\ref{fig:ModifiedMean},~\ref{fig:DSelfEnergy} and~\ref{fig:D0stSelfEnergy}. Blue (red) circles show the mass of $\bar{D}$ ($\bar{D}_0^*$) meson at each densities. Filled circles are the results with all one-loop corrections. Open circles show the masses of $\bar{D}$ and $\bar{D}_0^*$ mesons with only mean field $\sigma_0^*$ plotted as a reference. The dashed lines refer the mass of $\bar{D}$ and $\bar{D}_0^*$ mesons in the vacuum.}
\label{fig:MassOfDWithAll}
\end{figure}

Lastly, we summarize the density dependence of mass of $\bar{D}$ and $\bar{D}_0^*$ mesons with all one-loop corrections in Figs.~\ref{fig:ModifiedMean},~\ref{fig:DSelfEnergy} and~\ref{fig:D0stSelfEnergy}. The mass of $\bar{D}$ meson is defined by the solution of $q_0^2-\hat{m}_{\bar{D}}^2-{\rm Re}\tilde{\Sigma}_{\bar{D}}(q_0,\vec{0})=0$. The mass of $\bar{D}_0^*$ meson is defined by the value of $q_0$ at which the maximum of the first peak in the spectral function is realized. Density dependence of these masses are shown in Fig.~\ref{fig:MassOfDWithAll}.
Blue (red) circles are the mass of $\bar{D}$ ($\bar{D}_0^*$) meson at each density. Filled circles indicate their masses with all one-loop corrections in Figs.~\ref{fig:ModifiedMean},~\ref{fig:DSelfEnergy} and~\ref{fig:D0stSelfEnergy}, while open circles indicate their masses with mean field $\sigma_0^*$. By comparing Fig.~\ref{fig:MassOfDWithAll} 
with Fig.~\ref{fig:MassOfDWithModifiedMean} where only Hartree-type corrections are included,  we see that the Fock-type terms in Fig.~\ref{fig:DSelfEnergy} push 
down the mass of $\bar{D}$ meson while those in Fig.~\ref{fig:D0stSelfEnergy} 
push up the mass of $\bar{D}_0^*$ meson. 
In other word, Hartree-type and Fock-type diagrams generate mass modifications in opposite direction.  As a result, the density dependence of the masses of $\bar{D}$ and $\bar{D}_0^*$ mesons with all one-loop contributions is similar to the one with only mean field $\sigma_0^*$ included.

\begin{figure}[htbp]
\centering
\includegraphics*[scale=0.73]{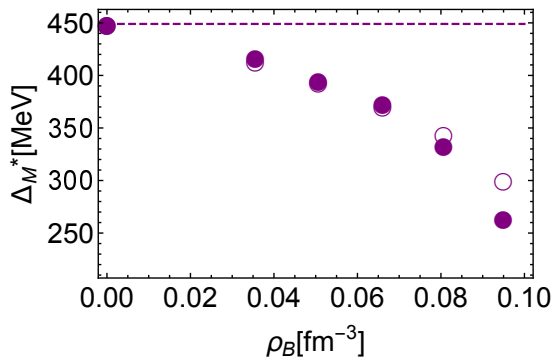}
\caption{(color online) Density dependence of mass difference between $\bar{D}_0^*$ meson and $\bar{D}$ meson with one-loop corrections shown in Figs.~\ref{fig:ModifiedMean},~\ref{fig:DSelfEnergy} and~\ref{fig:D0stSelfEnergy}.  Filled purple circles indicate the mass difference with all one-loop corrections, and open circles indicate the mass difference with mean field $\sigma_0^*$. The dashed line refers the mass difference between $\bar{D}_0^*$ and $\bar{D}$ mesons in the vacuum.}
\label{fig:DeltaMWithAll}
\end{figure}

Density dependence of mass difference between $\bar{D}_0^*$ and $\bar{D}$ mesons, $\Delta_m^*$, is shown in Fig.~\ref{fig:DeltaMWithAll}.
We can see that main contributions to the mass modifications to $\bar{D}$ and $\bar{D}_0^*$ mesons in nuclear matter is given by the mean field $\sigma_0^*$, and that the one-loop corrections in Figs.~\ref{fig:ModifiedMean},~\ref{fig:DSelfEnergy} and~\ref{fig:D0stSelfEnergy} are rather suppressed.

In obtaining Fig.~\ref{fig:MassOfDWithAll} and~\ref{fig:DeltaMWithAll}, we have employed the value of cutoff as $\Lambda=300$ MeV which is slightly higher than the scale of Fermi momentum. Cutoff dependence is also studied. When we choose it as $\Lambda=450$ MeV, the resulting masses of $\bar{D}$ and $\bar{D}_0^*$ mesons are changed $10$ MeV at most. Then cutoff dependence of our results is small.

In Ref.~\cite{Yasui:2012rw}, only $\bar{D}$ and $\bar{D}^*$ were included and it was pointed that the mass of $\bar{D}$ meson decreases by tens of MeV at normal nuclear matter density. In the present analysis, we include $\bar{D}$, $\bar{D}^*$, $\bar{D}_0^*$ and $\bar{D}_1$ mesons with chiral partner structure. Because of this structure, $\bar{D}$ meson mass is increased at mean field $\sigma_0^*$ level at density, and resultant mass which includes all one-loop corrections in Figs.~\ref{fig:ModifiedMean} and~\ref{fig:DSelfEnergy} at density is also larger than mass in the vacuum. As is seen from Fig.~\ref{fig:MassOfDWithModifiedMean} and Fig~\ref{fig:MassOfDWithAll}, Fock-type terms decease the $\bar{D}$ meson mass, and this behavior is consistent with result obtained in Ref.~\cite{Yasui:2012rw}.

\section{Conclusion}
\label{sec:Summary}
We study the mass and spectral functions for $\bar{D} (0^-)$ and $\bar{D}_0^*(0^+)$ mesons in nuclear matter in which the chiral symmetry is partially restored. $\bar{D}$ and $\bar{D}_0^*$ mesons are introduced to form the chiral partner structure. Then we focus on the modifications of $\bar{D}$ and $\bar{D}_0^*$ meson in nuclear matter since they are chiral partners to each other. Adopting the linear sigma model, we determine the ground state as the stationary point of $\sigma$ meson, and consider the fluctuations of pion, $\sigma$ meson and ``$\bar{D}$ mesons'' on such state perturbatively. Especially, in order to study the modifications of $\bar{D}$ meson and $\bar{D}_0^*$ meson in nuclear matter beyond the mean field level, we compute one-loop diagrams shown in Figs.~\ref{fig:ModifiedMean},~\ref{fig:DSelfEnergy} and~\ref{fig:D0stSelfEnergy}. In these diagrams, the two-point functions of $\sigma$ meson and pion should be resummed ones as shown in Fig.~\ref{fig:ResummedPropagator} to maintain the chiral symmetry.

When we take into account the mean field and Hartree-type terms in Fig.~\ref{fig:ModifiedMean}, the mass of $\bar{D}$ meson increases while that of $\bar{D}_0^*$ decreases more rapidly compared with mean field level. Then we can conclude that partial restoration of chiral symmetry at density is accelerated at one-loop level, and accordingly, mass difference between $\bar{D}_0^*$ and $\bar{D}$ mesons gets small compared with mean field level. When we take also Fock-type terms shown in Figs.~\ref{fig:DSelfEnergy} and~\ref{fig:D0stSelfEnergy}, however, the density dependence of mass of these mesons behave rather similar to mean field level. Then we can conclude that mean field effect is the dominant one when we study the mass of $\bar{D}$ and $\bar{D}_0^*$ mesons in nuclear matter. 

The spectral function for $\bar{D}_0^*$ and $\bar{D}$ meson channel are also studied. In the spectral function for $\bar{D}_0^*$ channel, three peaks are obtained as shown in Fig.~\ref{fig:SpectralEachKf}. The first peak from right essentially corresponds to the decay of $\bar{D}_0^*$ meson into $\bar{D}$ meson by emitting a pion, i.e., this peak represents the resonance of $\bar{D}_0^*$ meson state, and gradually shifts to lower energy. The height of this peak gets small as the density increases, which is caused by a collisional broadening. Second peak from right is induced by the threshold enhancement effect. This peak grows remarkably as the density increase and the peak position shifts to higher energy contrasts to the first peak from right. These changes are caused since the masses of $\bar{D}_0^*$ and $\bar{D}$ meson gets close as the density increased, which is a consequence of partial restoration of chiral symmetry. Particularly, the threshold enhancement can be a good probe to see the restoration of chiral symmetry, since this peak is exceedingly sharp. Third peak from right is the Landau damping which is the nuclear matter effect and this peak grows as the density increases.

There are several problems which are not covered in the present study. We do not take into account the effects of mean field of $\omega$ meson in this study. This effect can let the mass of $\bar{D}$ and $\bar{D}_0^*$ increase by a hundred MeV at most at normal nuclear matter density as studied in Ref.~\cite{Harada:2016uca}. The mass modifications to $\bar{D}$ and $\bar{D}_0^*$ mesons from $\omega$ contribution is the same, however, so that mass difference between $\bar{D}_0^*$ and $\bar{D}$ mesons are not changed. Besides, we do not include any charmed baryons such as $\bar{\Lambda}_c$$N$ loops. These loop corrections can be estimated as $\sim\frac{g_{\Lambda_c D N}^2}{2m_Nm_{\bar{\Lambda}_c}}\rho_B$, where $g_{\Lambda D N}$ is $\bar{\Lambda}_c$$\bar{D}$$N$ coupling. This correction can provide a few tens of MeV if $g_{\Lambda_c D N}$ is estimated as $g_{\Lambda_c D N}=10$, which is a natural choice of value of a hadron interaction. We need to include these corrections collectively, and we leave this work in the future publication.

In obtaining the spectral function in Fig.\ref{fig:SpectralEachKf}, we have treated the $\sigma$ meson as a stable state while the observed $\sigma$ meson has a width corresponding to the decay process of $\sigma\to \pi\pi$. When we include this effect, we expect that the first peak found in Fig.\ref{fig:SpectralEachKf} gets slightly broadened.

We construct a nuclear matter by the linear sigma model in this study. As is known well, such matter leads to the phase transition of chiral symmetry at lower than the normal nuclear matter density~\cite{Birse:1994cz}. However, we can show how can we adopt the fluctuations of $\sigma$ meson and pion perturbatively, and see the qualitative tendency of $\bar{D}$ mesons $\bar{D}_0^*$ meson in nuclear matter. We can easily apply the present method to more realistic matter such as obtained in Ref.~\cite{Motohiro:2015taa}.

 In the present analysis, we only consider the spectral function for $\bar{D}$ and $\bar{D}_0^*$ meson with $\vec{q}=\vec{0}$ for simplicity. In the experiment, however, it is expected it is difficult to measure the spectral functions at such particular kinematic region. Therefor we need to see them with nonzero momentum. We leave this work in the future publication.


\acknowledgments

This work is supported partly by the Grant-in-Aid for Scientific Research (Grant No.~25247036 and No.~15K17641) from Japan Society for the Promotion of Science (JSPS) (S.~Y.), and by the JSPS
Grant-in-Aid for Scientific Research (C) No. 16K05345 (M.~H.).


\appendix

\section{Calculation of the two-point vertex function $\tilde{\Gamma}_\sigma^{*(2)}(q_0,\vec{q})$ and $\tilde{\Gamma}_\pi^{*(2)}(q_0,\vec{q})$ in nuclear matter}
\label{sec:OneLoopApp}
Here, we shall show the calculation of the two-point vertex functions for pion and $\sigma$ meson $\tilde{\Gamma}^{*(2)}_\pi(q_0,\vec{q})$ ($\tilde{\Gamma}^{*(2)}_{\pi, ab}(q_0,\vec{q})=\delta^{ab}\tilde{\Gamma}^{*(2)}_\pi(q_0,\vec{q})$) and $\tilde{\Gamma}^{*(2)}_{\sigma}(q_0,\vec{q})$. These are provided by Eqs.~(\ref{TPPi}) and~(\ref{TPSigma}), and again we show explicitly here:
\begin{widetext}
\begin{eqnarray}
\tilde{\Gamma}^{*(2)}_\pi(q_0,\vec{q})&=&q^2-(m_0^2+\lambda\sigma_0^{*2})-2i g^2 \int\frac{\tilde{d}^4k}{(2\pi)^4}{\rm tr}\left[i\gamma_5\tilde{G}_N(k_0,\vec{k})i\gamma_5\tilde{G}_N(k_0-q_0,\vec{k}-\vec{q})\right]\nonumber\\
&\equiv&q^2-m_\pi^{*2}-i\tilde{\Sigma}^{*}_\pi(q_0,\vec{q})\ , \label{TwoPointPi}
\end{eqnarray}
\begin{eqnarray}
\tilde{\Gamma}^{*(2)}_\sigma(q_0,\vec{q})&=&q^2-(m_0^2+3\lambda\sigma_0^{*2})-2i g^2 \int\frac{\tilde{d}^4k}{(2\pi)^4}{\rm tr}\left[\tilde{G}_N(k_0,\vec{k})\tilde{G}_N(k_0-q_0,\vec{k}-\vec{q})\right] \nonumber\\
&\equiv&q^2-m_\sigma^{*2}-i\tilde{\Sigma}^{*}_\sigma(q_0,\vec{q})  \ .
\end{eqnarray}
\end{widetext}
$m_\pi^{*2}=m_0^2+\lambda\sigma_0^{*2}$ and $m_\sigma^{*2}=m_0^2+3\lambda\sigma_0^{*2}$ are the ``bare'' mass which are not identical to the on-shell mass. ``$\int\tilde{d}^4k/(2\pi)^4$'' stands for the momentum integration which depends on Fermi momentum $k_F$ explicitly defined in Eq.~(\ref{TildeInt}) as done in the gap equation~(\ref{GapDensDep}) in order to maintain the chiral symmetry. In fact, in the limit of $\epsilon\to0$, we can see $\left.\tilde{\Gamma}_\pi^{*(2)}(q_0,\vec{q})\right|_{q\to0}=0$ by inserting the gap equation~(\ref{GapDensDep}) as will be shown. $\tilde{G}_N(k_0,\vec{k})$ is the in-medium propagator of nucleon which carries momentum $k^\mu = (k_0,\vec{k})$ provided by Eq.~(\ref{InMediumPropagator}).
Note that the isospin factor for $\tilde{\Sigma}_{\pi}^{*}(q_0,\vec{q})$ is $2$ since ${\rm Tr}[\tau^a\tau^b] = 2\delta^{ab}$. The Feynman diagrams of self-energies $\tilde{\Sigma}_\pi^*(q_0,\vec{q})$ and $\tilde{\Sigma}_\sigma(q_0,\vec{q})$ are displayed in Figs.~\ref{fig:SelfEnergies}. The inverse of two-point vertex functions, i.e., the two-point functions of pion and $\sigma$ meson are the ones which resum nucleon loops as shown in Fig.~\ref{fig:ResummedPropagator}.

In order to get the explicit expression of the spectral functions $\rho_\pi^*(q_0,\vec{q})$ and $\rho_\sigma^*(q_0,\vec{q})$ in Eq.~(\ref{RenormalizedSpectral}), it is convenient to calculate the imaginary part and real part of retarded self-energy in Fig.~\ref{fig:SelfEnergies} separately. The imaginary part is easily obtained as
\begin{widetext}
\begin{eqnarray}
{\rm Im}\tilde{\Sigma}^{*R}_\pi(q_0,\vec{q}) &=&4 \pi g^2q^2\int\frac{d^3k}{(2\pi)^3}\frac{1}{4E_1E_2}\theta(k_F-|\vec{k}|)\delta(q_0-E_1-E_2) \nonumber\\
&-&  4 \pi g^2q^2\int\frac{d^3k}{(2\pi)^3}\frac{1}{4E_1E_2}\theta(k_F-|\vec{k}|)\delta(q_0+E_1+E_2) \nonumber\\
&-&  4 \pi  g^2q^2\int\frac{d^3k}{(2\pi)^3}\frac{1}{4E_1E_2}\theta(k_F-|\vec{k}|)\delta(q_0-E_1+E_2) \nonumber\\
&+&4 \pi  g^2q^2\int\frac{d^3k}{(2\pi)^3}\frac{1}{4E_1E_2}\theta(k_F-|\vec{k}|)\delta(q_0+E_1-E_2)  \ , \label{ImPiMatter}
\end{eqnarray}
and 
\begin{eqnarray}
{\rm Im}\tilde{\Sigma}^{*R}_\sigma(q_0,\vec{q}) &=& 4\pi g^2(q^2-4m_N^{*2})\int\frac{d^3k}{(2\pi)^3}\frac{1}{4E_1E_2}\theta(k_F-|\vec{k}|)\delta(q_0-E_1-E_2) \nonumber\\
&-& 4 \pi g^2(q^2-4m_N^{*2})\int\frac{d^3k}{(2\pi)^3}\frac{1}{4E_1E_2}\theta(k_F-|\vec{k}|)\delta(q_0+E_1+E_2) \nonumber\\
&-&  4 \pi  g^2(q^2-4m_N^{*2})\int\frac{d^3k}{(2\pi)^3}\frac{1}{4E_1E_2}\theta(k_F-|\vec{k}|)\delta(q_0-E_1+E_2) \nonumber\\
&+& 4 \pi  g^2(q^2-4m_N^{*2})\int\frac{d^3k}{(2\pi)^3}\frac{1}{4E_1E_2}\theta(k_F-|\vec{k}|)\delta(q_0+E_1-E_2)  \ ,\label{ImSigmaMatter}
\end{eqnarray}
\end{widetext}
where $\epsilon(q_0)$ is the sign function which is defined as $\epsilon(q_0) = +1(-1)$ for $q_0>0(q_0<0)$, and $\theta(x)$ is the Heaviside step function. $E_1$, $E_2$ are defined as 
\begin{eqnarray}
E_1 &=& \sqrt{|\vec{k}|^2+m_N^{*2}} \nonumber\\
E_2 &=& \sqrt{|\vec{k}-\vec{q}|^2+m_N^{*2}}\ .
\end{eqnarray}
The first and second terms in Eqs.~(\ref{ImPiMatter}) and~(\ref{ImSigmaMatter}) correspond to the pair creation or annihilation of nucleon and anti-nucleon and these terms do not vanish only when $q^2-4m_N^{*2}>0$ is satisfied. The third and fourth terms correspond to the Landau damping, and these can contribute only when $q^2<0$. The real parts of the retarded self-energies in Fig~\ref{fig:SelfEnergies} can be obtained straightforwardly via the dispersion relation~(\ref{Dispersion1}), because these contributions do not suffer from any divergences due to the cut-off of Fermi momentum $k_F$ in the three momentum integral. 
Alternatively, the real parts are directly computed as
\begin{widetext}
\begin{eqnarray}
{\rm Re}\tilde{\Sigma}_\pi^{*R}(q_0,\vec{q}) &=&  \frac{g^2}{2\pi^2} \int_0^{k_F}d|\vec{k}|\frac{|\vec{k}|^2}{E_1}\left\{4-\frac{q_0^2-|\vec{q}|^2}{2|\vec{k}||\vec{q}|}{\rm ln}\left|\frac{q_0^2-|\vec{q}|^2+2|\vec{k}||\vec{q}|+2q_0E_1}{q_0^2-|\vec{q}|^2-2|\vec{k}||\vec{q}|+2q_0E_1}\frac{q_0^2-|\vec{q}|^2+2|\vec{k}||\vec{q}|-2q_0E_1}{q_0^2-|\vec{q}|^2-2|\vec{k}||\vec{q}|-2q_0E_1}\right|\right\} \nonumber\\  \label{PiReal}
\end{eqnarray}
and
\begin{eqnarray}
{\rm Re}\tilde{\Sigma}^{*R}_\sigma(q_0,\vec{q}) &=& \frac{ g^2 }{ 2\pi^2} \int_0^{k_F}d|\vec{k}|\frac{|\vec{k}|^2}{E_1}\left\{4-\frac{q_0^2-|\vec{q}|^2-4m_N^{*2}}{2|\vec{k}||\vec{q}|}{\rm ln}\left|\frac{q_0^2-|\vec{q}|^2+2|\vec{k}||\vec{q}|+2q_0E_1}{q_0^2-|\vec{q}|^2-2|\vec{k}||\vec{q}|+2q_0E_1}\frac{q_0^2-|\vec{q}|^2+2|\vec{k}||\vec{q}|-2q_0E_1}{q_0^2-|\vec{q}|^2-2|\vec{k}||\vec{q}|-2q_0E_1}\right|\right\} \ .\nonumber\\  \label{SigmaReal}
\end{eqnarray}
\end{widetext}
From these ${\rm Im}\tilde{\Sigma}_{\pi(\sigma)}^{*R}(q_0,\vec{q})$ and ${\rm Re}\tilde{\Sigma}_{\pi(\sigma)}^{*R}(q_0,\vec{q})$, we can find explicit form of spectral functions in Eq.~(\ref{RenormalizedSpectral}).

Finally, let us check the chiral symmetry by seeing a massless solution in $\tilde{\Gamma}_\pi^{*(2)}(q_0,\vec{0})=0$ in the chiral limit as mentioned before. In the limit of $\vec{q}\to\vec0$, real part of the retarded self-energy for pion propagator~(\ref{PiReal}) is reduced to
\begin{eqnarray}
{\rm Re}\tilde{\Sigma}_\pi^{*R}(q_0,\vec{0}) &\to&  \frac{2 g^2}{\pi^2}\int_0^{k_F}d|\vec{k}|\frac{|\vec{k}|^2}{E_1} \nonumber\\
&& -q_0^2  \frac{ 2 g^2 }{\pi^2}  \int_0^{k_F}d|\vec{k}|\frac{| \vec{k}|^2}{E_1}\frac{1}{q_0^2-4E_1^{2}} \ ,\nonumber\\
\end{eqnarray}
and imaginary part vanishes in this domain. Then two-point vertex function of pion in Eq.~(\ref{TwoPointPi}) together with gap equation~(\ref{GapDensDep}) is reduced to
\begin{eqnarray}
\tilde{\Gamma}_\pi^{*(2)}(q_0,\vec{0}) &\to&  q_0^2\left[1+  \frac{ 2 g^2 }{\pi^2}\int_0^{k_F}d|\vec{k}|\frac{| \vec{k}|^2}{E_1}\frac{1}{q_0^2-4E_1^{2}}\right] \nonumber\\
&& -\frac{\epsilon}{\sigma_0^*}\ ,
\end{eqnarray}
which reads the solution $\tilde{\Gamma}_\pi^{*(2)}(0,\vec{0})=0$, i.e., $\tilde{m}_\pi^2=0$ in the chiral limit $\epsilon \to 0$. Therefor, when we calculated the fluctuations around the mean field, we have to use the resummed propagator which is shown in Fig.~\ref{fig:ResummedPropagator}.


\section{Derivation of the spectral function in Eq.~(\ref{RenormalizedSpectral})}
\label{sec:Dispersion}
Here, we derive the spectral function in Eq.~(\ref{RenormalizedSpectral}). The two-point function of a spin-0 particle in the coordinate space $G(x_0,\vec{x}) = \langle{\rm T}\phi(x_0,\vec{x})\phi(0,\vec{0})\rangle$ is decomposed into two pieces as
\begin{eqnarray}
&&G(x_0,\vec{x})  \nonumber\\
&&\ \ \ \ \ \ \ = \theta(x_0)\langle\phi(x_0,\vec{x})\phi(0,\vec{0})\rangle + \theta(-x_0)\langle\phi(0,\vec{0})\phi(x_0,\vec{0})\rangle \nonumber\\
&&\ \ \ \ \ \ \  \equiv \theta(x_0)G^>(x_0,\vec{x}) + \theta(-x_0)G^<(x_0,\vec{x})\ , \label{Decomposition1}\nonumber\\
\end{eqnarray}
where we have defined the greater Green's function $G^>(x_0,\vec{x})$ and lesser Green's function $G^<(x_0,\vec{x})$ by
\begin{eqnarray}
G^>(x_0,\vec{x}) &\equiv& \langle{\phi}(x_0,\vec{x}){\phi}(0,\vec{0})\rangle \nonumber\\
G^<(x_0,\vec{x}) &\equiv& \langle{\phi}(0,\vec{0}){\phi}(x_0,\vec{x})\rangle \label{GLGreen}
\end{eqnarray}
for later uses. The retarded Green's function $G^R(x_0,\vec{x})$ is defined by
\begin{eqnarray}
G^R(x_0,\vec{x}) &\equiv& i\theta(x_0)\left( \langle{\phi}(x_0,\vec{x}){\phi}(0,\vec{0})\rangle - \langle{\phi}(0,\vec{0}){\phi}(x_0,\vec{x})\rangle\right)  \nonumber\\
&=& i\theta(x_0)\left(G^>(x_0,\vec{x})-G^<(x_0,\vec{x})\right)  \label{RetardedDef}\ .
\end{eqnarray}
By using 
\begin{eqnarray}
\theta(x_0) = i\int\frac{dk_0}{2\pi}\frac{{\rm e}^{-ik_0 t}}{k_0+i\epsilon}\ ,
\end{eqnarray}
the Fourier transformation of the retarded Green's function in Eq.~(\ref{RetardedDef}) takes
\begin{eqnarray}
\tilde{G}^R(q_0,\vec{q}) &=& -\int\frac{dk_0}{2\pi}\frac{\tilde{G}^>(k_0,\vec{k})-\tilde{G}^<(k_0,\vec{k})}{q_0-k_0+i\epsilon}\nonumber\\
&\equiv& -\int\frac{dk_0}{2\pi}\frac{\rho(k_0,\vec{q})}{q_0-k_0+i\epsilon}\nonumber\\
&=&-\int \frac{dk_0}{2\pi}{\rm P}\, \frac{\rho(k_0,\vec{q})}{q_0-k_0}+i\frac{1}{2}\rho(q_0,\vec{q})  \ , \label{FourierRetarded}
\end{eqnarray}
where $\tilde{G}^R(q_0,\vec{q})$, $\tilde{G}^>(q_0,\vec{q})$, $\tilde{G}^<(q_0,\vec{q})$ are the Fourier transformation of $G^R(x_0,\vec{x})$, $G^>(x_0,\vec{x})$, $G^<(x_0,\vec{x})$, respectively. ${\rm P}$ is the symbol of principal value integral. In the second line in Eq.~(\ref{FourierRetarded}), we have defined the spectral function $\rho(k_0,\vec{k})$ as
\begin{eqnarray}
\rho(q_0,\vec{q}) \equiv \tilde{G}^>(q_0,\vec{q})-\tilde{G}^<(q_0,\vec{q})\ .\label{Spectral}
\end{eqnarray}
Then according to Eq.~(\ref{FourierRetarded}), the spectral function of a spin-0 particle satisfies
\begin{eqnarray}
\rho(q_0,\vec{q}) = 2{\rm Im}\tilde{G}^R(q_0,\vec{q})\ .  \label{RhoRelation}
\end{eqnarray}
Eq.~(\ref{RhoRelation}) tells the spectral function is provided by the imaginary part of retarded Green's function. 

The self-energy $\Sigma(x_0,\vec{x})$ is also decomposed into two pieces as
\begin{eqnarray} 
\Sigma(x_0,\vec{x}) = \theta(x_0)\Sigma^>(x_0,\vec{x}) + \theta(-x_0)\Sigma^<(x_0,\vec{x})\ , \nonumber\\ \label{SigmaDec}
\end{eqnarray}
in the same manner as in Eq.~(\ref{Decomposition1}). When we defined the retarded self-energy in the coordinate space $\Sigma^R(x_0,\vec{x})$ by 
\begin{eqnarray}
\Sigma^R(x_0,\vec{x}) = i\theta(x_0)\left(\Sigma^>(x_0,\vec{x}) - \Sigma^<(x_0,\vec{x})\right)\ , \label{RetardedSigma}
\end{eqnarray}
the retarded Green's function in the momentum space $\tilde{G}^R(q_0,\vec{q})$ with resumming retarded self-energy is of the form
\begin{eqnarray}
\tilde{G}^R(q_0,\vec{q}) = -\frac{1}{q^2-m^2-\tilde{\Sigma}^R(q_0,\vec{q})}\ ,
\end{eqnarray}
where $m$ is the bare mass of scalar field ${\phi}(x_0,\vec{x})$ and $\tilde{\Sigma}^R(q_0,\vec{q})$ is the Fourier transformation of $\Sigma^R(x_0,\vec{x})$. Therefor, Eq.~(\ref{RhoRelation}) is expressed in terms of the retarded self-energy $\tilde{\Sigma}^R(q_0,\vec{q})$ as
\begin{eqnarray}
\rho(q_0,\vec{q}) = \frac{-2{\rm Im}\tilde{\Sigma}^R(q_0,\vec{q})}{\left[ q^2-m^2-{\rm Re}\tilde{\Sigma}^R(q_0,\vec{q})\right]^2+\left[{\rm Im}\tilde{\Sigma}^R(q_0,\vec{q})\right]^2} \ .\nonumber\\
\end{eqnarray}

Furthermore, according to Eqs.~(\ref{Spectral}) and~(\ref{RhoRelation}), we obtain a relation among ${\rm Im}\tilde{G}^R(q)$, $\tilde{G}^>(q_0,\vec{q})$ and $\tilde{G}^<(q_0,\vec{q})$ as
\begin{eqnarray}
{\rm Im}\tilde{G}^R(q_0,\vec{q}) = \frac{1}{2}\left(\tilde{G}^>(q_0,\vec{q})-\tilde{G}^<(q_0,\vec{q})\right)\ . \label{ImRelation}
\end{eqnarray}
Since we only have used the property of Heaviside's step function $\theta(x_0)$ and $\theta(-x_0)$ in obtaining Eq.~(\ref{ImRelation}), a similar relation holds for the self energy:
\begin{eqnarray}
{\rm Im}\tilde{\Sigma}^R(q_0,\vec{q}) = \frac{1}{2}\left(\tilde{\Sigma}^>(q_0,\vec{q})-\tilde{\Sigma}^<(q_0,\vec{q})\right)\ , \label{ImRelationSigma}
\end{eqnarray}
where $\tilde{\Sigma}^>(q_0,\vec{q})$ and $\tilde{\Sigma}^<(q_0,\vec{q})$ is defined by the Fourier transformation of $\Sigma^>(x_0,\vec{x})$ and $\Sigma^<(x_0,\vec{x})$, respectively.


\section{The dispersion relation}
\label{sec:SubDispersion}
Here, we provide a derivation of the dispersion relation which is so-called Kramers-Kronig relation. This relation connects the imaginary part of retarded self-energy ${\rm Im}\tilde{\Sigma}^R(q_0,\vec{q})$ and the real part of retarded self-energy ${\rm Re}\tilde{\Sigma}^R(q_0,\vec{q})$. 

The retarded self-energy $\Sigma^R(x_0,\vec{x})$ in the coordinate space is defined by~(\ref{RetardedSigma}), then $\Sigma^R(x_0,\vec{x})$ satisfies 
\begin{eqnarray}
\Sigma^R(x_0,\vec{x}) = \theta(x_0)\Sigma^R(x_0,\vec{x})
\end{eqnarray}
due to the causality. Accordingly, the Fourier transformation of $\Sigma^R(x_0,\vec{x})$ satisfies
\begin{eqnarray}
\tilde{\Sigma}^R(q_0,\vec{q}) &=& i\int\frac{dk_0}{2\pi}\frac{\tilde{\Sigma}^R(k_0,\vec{q})}{q_0-k_0+i\epsilon} \nonumber\\
&=& i{\rm P}\, \int\frac{dk_0}{2\pi}\frac{\tilde{\Sigma}^R(k_0,\vec{q})}{q_0-k_0}+\frac{1}{2}\tilde{\Sigma}_R(q_0,\vec{q}) \ ,
\end{eqnarray}
which leads to the dispersion relation
\begin{eqnarray}
{\rm Re}\tilde{\Sigma}^R(q_0,\vec{q}) &=& \frac{1}{\pi}{\rm P}\, \int_{-\infty}^{\infty} dz\frac{{\rm Im}\tilde{\Sigma}^R(z,\vec{q})}{z-q_0} \ .\nonumber\\
&=& \frac{1}{\pi}{\rm P}\, \int_{0}^{\infty} dz^2\frac{{\rm Im}\tilde{\Sigma}^R(z,\vec{q})}{z^2-q_0^2} 
\label{Dispersion1}
\end{eqnarray}
In obtaining the second line in Eq.~(\ref{Dispersion1}), we have used a relation of ${\rm Im}\tilde{\Sigma}^{R}(q_0,\vec{q}) =\epsilon(q_0){\rm Im}\left(i\tilde{\Sigma}(q_0,\vec{q})\right)$ where $\tilde{\Sigma}(q_0,\vec{q})$ is defined by the Fourier transformation of $\Sigma(x_0,\vec{x}).$ This is a consequence of the charge conjugation property which holds since we include the matter
effect through the nucleon-hole one-loop contribution to the $\sigma$ and $\pi$ propagator.

When ${\rm Re}\tilde{\Sigma}^R(q_0,\vec{q})$ is suffered by some logarithmic UV divergences, ~(\ref{Dispersion1}) should be modified by the following subtracted dispersion relation to regularize it:
\begin{widetext}
\begin{eqnarray}
{\rm Re}\tilde{\Sigma}^R(q_0,\vec{q}) &=&  \frac{1}{\pi}{\rm P}\, \int_{-\infty}^{\infty} dz^2\frac{{\rm Im}\tilde{\Sigma}^R(z,\vec{q})}{z^2-q_0^2} \nonumber\\ \nonumber\\
&\to&  \frac{1}{\pi}{\rm P}\, \int_{0}^{\infty} dz^2\left(\frac{{\rm Im}\tilde{\Sigma}^R(z,\vec{q})}{z^2-q_0^2} -\frac{{\rm Im}\tilde{\Sigma}^R(z,\vec{q})}{z^2-|\vec{q}|^2-m^2}\right)\nonumber\\ 
&=&  \frac{q^2-m^2}{\pi}{\rm P}\, \int_0^{\infty} dz^2\frac{{\rm Im}\tilde{\Sigma}^R(z,\vec{q})}{(z^2-q_0^2)(z^2-|\vec{q}|^2-m^2)}\ , \nonumber\\ \label{SubDispersion}
\end{eqnarray}
\end{widetext}
where $m$ is the renormalized mass of scalar field since ${\rm Re}\tilde{\Sigma}^R(q_0,\vec{q})$ satisfies ${\rm Re}\tilde{\Sigma}^R(\sqrt{|\vec{q}|^2+m^2},\vec{q})=0$.

\section{Properties of $\tilde{G}^>(q_0,\vec{q})$ and $\tilde{G}^<(q_0,\vec{q})$}
\label{sec:GL}
Here, we shall find the explicit form of $\tilde{G}^>(q_0,\vec{q})$ and $\tilde{G}^<(q_0,\vec{q})$ at zero temperature. When a scalar field $\phi(x_0,\vec{x})$ is a free particle, $\phi(x_0,\vec{x})$ is expanded in terms of the creation and annihilation operator $a_{q}^\dagger$ and $a_{q}$ as
\begin{eqnarray}
\phi(x_0,\vec{x}) = \int\frac{d^3q}{(2\pi)^32\epsilon_q}\left\{a_{q}\, {\rm e}^{-iq\cdot x}+a^\dagger_{q}\, {\rm e}^{iq\cdot x}\right\}\ ,
\end{eqnarray}
where $\epsilon_q\equiv \sqrt{|\vec{q}|^2+m^2}$ and $q\cdot x= \epsilon_qx_0-\vec{q}\cdot\vec{x}$. $a_{q}$ and $a_{q}^\dagger$ satisfies the following commutation relation
\begin{eqnarray}
[\, a_{q}, a^\dagger_{p}\, ] = (2\pi)^32\epsilon_q\delta^3(\vec{q}-\vec{p})\ ,
\end{eqnarray}
and the vacuum $|0\rangle$ is defined by $a_q|0\rangle=0$. Then from Eq.~(\ref{GLGreen}), we can easily find 
\begin{eqnarray}
\tilde{G}_{\rm free}^>(x_0,\vec{x}) &=& \left.\langle0|\phi(x_0,\vec{x})\phi(y_0,\vec{y})|0\rangle\right|_{y_0=0,\vec{y}=\vec{0}} \nonumber\\
&=& \int \frac{d^3q}{(2\pi)^3}\frac{1}{2\epsilon_q}{\rm e}^{-i\epsilon_qx_0+i\vec{q}\cdot\vec{x}} \nonumber\\
&=&  \int \frac{d^4q}{(2\pi)^4}\frac{2\pi}{2\epsilon_q}\delta(q_0-\epsilon_q){\rm e}^{-iq\cdot x}\ , 
\end{eqnarray}
and similarly 
\begin{eqnarray}
\tilde{G}_{\rm free}^<(x_0,\vec{x}) &=& \left.\langle0|\phi(y_0,\vec{y})\phi(x_0,\vec{x})|0\rangle\right|_{y_0=0,\vec{y}=\vec{0}} \nonumber\\
&=& \int \frac{d^3q}{(2\pi)^3}\frac{1}{2\epsilon_q}{\rm e}^{+i\epsilon_qx_0-i\vec{q}\cdot\vec{x}} \nonumber\\
&=&  \int \frac{d^4q}{(2\pi)^4}\frac{2\pi}{2\epsilon_q}\delta(q_0+\epsilon_q){\rm e}^{-iq\cdot x}\ , 
\end{eqnarray}
which reads
\begin{eqnarray}
\tilde{G}_{\rm free}^>(q_0,\vec{q})  &=& \frac{2\pi}{2\epsilon_q}\delta(q_0-\epsilon_q) \nonumber\\
\tilde{G}_{\rm free}^<(q_0,\vec{q}) &=&\frac{2\pi}{2\epsilon_q}\delta(q_0+\epsilon_q) \ .\label{GLVac}
\end{eqnarray}
Eq.~(\ref{GLVac}) yields the spectral function $\rho_{\rm free}(q_0,\vec{q})$ which is defined by Eq.~(\ref{Spectral}) as
\begin{eqnarray}
\rho_{\rm free}(q_0,\vec{q}) &=& \frac{2\pi}{2\epsilon_q}\delta(q_0-\epsilon_q)-\frac{2\pi}{2\epsilon_q}\delta(q_0+\epsilon_q) \nonumber\\
&=& 2\pi\epsilon(q_0)\delta(q^2-m^2)\ , \label{RhoVac}
\end{eqnarray}
where $\epsilon(q_0)$ is the sign function defined by $\epsilon(q_0)=+1(-1)$ for $q_0>0 \ (q_0<0)$. Namely, $\tilde{G}_{\rm free}^>(q_0,\vec{q})$ and $\tilde{G}_{\rm free}^<(q_0,\vec{q})$ in Eq.~(\ref{GLVac}) are rewritten in terms of $\rho_{\rm free}(q_0,\vec{q})$ as
\begin{eqnarray}
\tilde{G}_{\rm free}^>(q_0,\vec{q})  &=& \theta(q_0)\rho_{\rm free}(q_0,\vec{q})\nonumber\\
\tilde{G}_{\rm free}^<(q_0,\vec{q}) &=&-\theta(-q_0)\rho_{\rm free}(q_0,\vec{q})\ . \label{GLVac2}
\end{eqnarray}
Eq.~(\ref{GLVac2}) is the explicit form of $\tilde{G}^>(q_0,\vec{q})$ and $\tilde{G}^<(q_0,\vec{q})$ for a free particle. When the spin-0 particle is not a free particle, the spectral function is replaced by general one, but Eq.~(\ref{GLVac2}) still holds:
\begin{eqnarray}
\tilde{G}^>(q_0,\vec{q})  &=& \theta(q_0)\rho(q_0,\vec{q})\nonumber\\
\tilde{G}^<(q_0,\vec{q}) &=&-\theta(-q_0)\rho(q_0,\vec{q})\ . \label{GLRel}
\end{eqnarray}
Note that Eq.~(\ref{GLRel}) is realized only when the distribution function for the spin-0 particle is not disturbed by medium~\cite{TFT}.


\begin{widetext}
\section{Calculations of self-energies in Figs.~\ref{fig:DSelfEnergy} and~\ref{fig:D0stSelfEnergy}}
\label{sec:ImaginaryParts}
Here, we show calculations of all self-energies in Fig.~\ref{fig:DSelfEnergy} and~\ref{fig:D0stSelfEnergy}. The imaginary parts are calculated in the same way as done in Eq.~(\ref{ImSigmaComp}). ${\rm Im}\tilde{\Sigma}_{\bar{D}(1a)}^{*R}(q_0,\vec{q})$, ${\rm Im}\tilde{\Sigma}_{\bar{D}(1b)}^{*R}(q_0,\vec{q})$, ${\rm Im}\tilde{\Sigma}_{\bar{D}(1c)}^{*R}(q_0,\vec{q})$ and ${\rm Im}\tilde{\Sigma}_{\bar{D}(1d)}^{*R}(q_0,\vec{q})$ are
\begin{eqnarray}
{\rm Im}\tilde{\Sigma}_{\bar{D}(1a)}^{*R}(q_0,\vec{q}) &=& - \frac{1}{2}m^2G_\pi^2\int\frac{d^4k}{(2\pi)^4}\left(F(\vec{k};\Lambda)\right)^2\theta(k_0)\rho_{\bar{D}}(k_0,\vec{q})\theta(q_0-k_0)\rho_{\sigma}^*(q_0-k_0,\vec{q}-\vec{k}) \nonumber\\
&&  +\frac{1}{2}m^2G_\pi^2\int\frac{d^4k}{(2\pi)^4}\left(F(\vec{k};\Lambda)\right)^2\theta(-k_0)\rho_{\bar{D}}(k_0,\vec{k})\theta(-q_0+k_0)\rho_\sigma^*(q_0-k_0,\vec{q}-\vec{k})\nonumber\\
&=&-\frac{1}{2}m^2G_\pi^2\int\frac{d^3k}{(2\pi)^3}\left(F(\vec{k};\Lambda)\right)^2 \nonumber\\
&&\times \frac{1}{2E_k^{\bar{D}}}\left\{\theta(q_0-E_k^{\bar{D}})\rho^*_\sigma(q_0-E_k^{\bar{D}},\vec{q}-\vec{k})+\theta(-E_k^{\bar{D}}-q_0)\rho^*_\sigma(q_0+E_k^{\bar{D}},\vec{q}-\vec{k})\right\}  \\ \nonumber\\
{\rm Im}\tilde{\Sigma}_{\bar{D}(1b)}^{*R}(q_0,\vec{q})
 &=& -\frac{3}{2}m^2G_\pi^2\int\frac{d^4k}{(2\pi)^4}\left(F(\vec{k};\Lambda)\right)^2\theta(k_0)\rho_{\bar{D}_0^*}(k_0,\vec{q})\theta(q_0-k_0)\rho_{\pi}^*(q_0-k_0,\vec{q}-\vec{k})\nonumber\\
 && +\frac{3}{2}m^2G_\pi^2\int\frac{d^4k}{(2\pi)^4}\left(F(\vec{k};\Lambda)\right)^2\theta(-k_0)\rho_{\bar{D}_0^*}(k_0,\vec{k})\theta(-q_0+k_0)\rho_\pi^*(q_0-k_0,\vec{q}-\vec{k})\nonumber\\
&=&-\frac{3}{2}m^2G_\pi^2\int\frac{d^3k}{(2\pi)^3}\left(F(\vec{k};\Lambda)\right)^2 \nonumber\\
&&\times\frac{1}{2E_k^{\bar{D}_0^*}}\left\{\theta(q_0-E_k^{\bar{D}_0^*})\rho^*_\pi(q_0-E_k^{\bar{D}_0^*},\vec{q}-\vec{k})+\theta(-E_k^{\bar{D}_0^*}-q_0)\rho^*_\pi(q_0+E_k^{\bar{D}_0^*},\vec{q}-\vec{k})\right\}  \\ \nonumber\\ 
{\rm Im}\tilde{\Sigma}_{\bar{D}(1c)}^{*R}(q_0,\vec{q}) &=& \frac{3}{2}\left(g_A\frac{m}{f_\pi}\right)^2\int\frac{d^4k}{(2\pi)^4}\left(F(\vec{k};\Lambda)\right)^2(q-k)^\mu (q-k)^\nu P^{\bar{D}^*}_{\mu\nu}\theta(k_0)\rho_{\bar{D}^*}(k_0,\vec{q})\theta(q_0-k_0)\rho_{\pi}^*(q_0-k_0,\vec{q}-\vec{k}) \nonumber\\
&& - \frac{3}{2}\left(g_A\frac{m}{f_\pi}\right)^2\int\frac{d^4k}{(2\pi)^4}\left(F(\vec{k};\Lambda)\right)^2(q-k)^\mu (q-k)^\nu P^{\bar{D}^*}_{\mu\nu}\theta(-k_0)\rho_{\bar{D}^*}(k_0,\vec{q})\theta(-q_0+k_0)\rho_{\pi}^*(q_0-k_0,\vec{q}-\vec{k}) \nonumber\\
&=&-\frac{3}{2}q_0^2\left(g_A\frac{m}{f_\pi}\right)^2\int\frac{d^3k}{(2\pi)^3}\left(F(\vec{k};\Lambda)\right)^2 \nonumber\\
&&\times\frac{1}{2E_k^{\bar{D}^*}} \frac{|\vec{k}|^2}{m_{\bar{D}^*}^2}\left\{\theta(q_0-E_k^{\bar{D}^*})\rho^*_\pi(q_0-E_k^{\bar{D}^*},\vec{q}-\vec{k})+\theta(-E_k^{\bar{D}^*}-q_0)\rho^*_\pi(q_0+E_k^{\bar{D}^*},\vec{q}-\vec{k})\right\}   \\ \nonumber\\ 
{\rm Im}\tilde{\Sigma}_{\bar{D}(1d)}^{*R}(q_0,\vec{q}) &=&\frac{1}{2}\left(g_A\frac{m}{f_\pi}\right)^2\int\frac{d^4k}{(2\pi)^4}\left(F(\vec{k};\Lambda)\right)^2(q-k)^\mu (q-k)^\nu P^{\bar{D}_1}_{\mu\nu}\theta(k_0)\rho_{\bar{D}_1}(k_0,\vec{q})\theta(q_0-k_0)\rho_{\sigma}^*(q_0-k_0,\vec{q}-\vec{k}) \nonumber\\
&& - \frac{1}{2}\left(g_A\frac{m}{f_\pi}\right)^2\int\frac{d^4k}{(2\pi)^4}\left(F(\vec{k};\Lambda)\right)^2(q-k)^\mu (q-k)^\nu P^{\bar{D}_1}_{\mu\nu}\theta(-k_0)\rho_{\bar{D}_1}(k_0,\vec{q})\theta(-q_0+k_0)\rho_{\sigma}^*(q_0-k_0,\vec{q}-\vec{k}) \nonumber\\
&=&-\frac{1}{2}q_0^2\left(g_A\frac{m}{f_\pi}\right)^2\int\frac{d^3k}{(2\pi)^3}\left(F(\vec{k};\Lambda)\right)^2 \nonumber\\
&&\times\frac{1}{2E_k^{\bar{D}_1}} \frac{|\vec{k}|^2}{m_{\bar{D}_1}^2}\left\{\theta(q_0-E_k^{\bar{D}_1})\rho^*_\sigma(q_0-E_k^{\bar{D}_1},\vec{q}-\vec{k})+\theta(-E_k^{\bar{D}_1}-q_0)\rho^*_\sigma(q_0+E_k^{\bar{D}_1},\vec{q}-\vec{k})\right\}    \ .  
\end{eqnarray}
${\rm Im}\tilde{\Sigma}_{\bar{D}_0^*(2a)}^{*R}(q_0,\vec{q})$, ${\rm Im}\tilde{\Sigma}_{\bar{D}_0^*(2b)}^{*R}(q_0,\vec{q})$, ${\rm Im}\tilde{\Sigma}_{\bar{D}_0^*(2c)}^{*R}(q_0,\vec{q})$ and ${\rm Im}\tilde{\Sigma}_{\bar{D}_0^*(2d)}^{*R}(q_0,\vec{q})$ are
\begin{eqnarray}
{\rm Im}\tilde{\Sigma}_{\bar{D}_0^*(2a)}^{*R}(q_0,\vec{q}) &=& - \frac{3}{2}m^2G_\pi^2\int\frac{d^4k}{(2\pi)^4}\left(F(\vec{k};\Lambda)\right)^2\theta(k_0)\rho_{\bar{D}}(k_0,\vec{q})\theta(q_0-k_0)\rho_{\pi}^*(q_0-k_0,\vec{q}-\vec{k}) \nonumber\\
&&  +\frac{3}{2}m^2G_\pi^2\int\frac{d^4k}{(2\pi)^4}\left(F(\vec{k};\Lambda)\right)^2\theta(-k_0)\rho_{\bar{D}}(k_0,\vec{k})\theta(-q_0+k_0)\rho_\pi^*(q_0-k_0,\vec{q}-\vec{k})\nonumber\\
&=&-\frac{3}{2}m^2G_\pi^2\int\frac{d^3k}{(2\pi)^3}\left(F(\vec{k};\Lambda)\right)^2 \nonumber\\
&&\times \frac{1}{2E_k^{\bar{D}}}\left\{\theta(q_0-E_k^{\bar{D}})\rho^*_\pi(q_0-E_k^{\bar{D}},\vec{q}-\vec{k})+\theta(-E_k^{\bar{D}}-q_0)\rho^*_\pi(q_0+E_k^{\bar{D}},\vec{q}-\vec{k})\right\}  \\ \nonumber\\
{\rm Im}\tilde{\Sigma}_{\bar{D}_0^*(2b)}^{*R}(q_0,\vec{q}) &=& - \frac{1}{2}m^2G_\pi^2\int\frac{d^4k}{(2\pi)^4}\left(F(\vec{k};\Lambda)\right)^2\theta(k_0)\rho_{\bar{D}_0^*}(k_0,\vec{q})\theta(q_0-k_0)\rho_{\sigma}^*(q_0-k_0,\vec{q}-\vec{k}) \nonumber\\
&&  +\frac{1}{2}m^2G_\pi^2\int\frac{d^4k}{(2\pi)^4}\left(F(\vec{k};\Lambda)\right)^2\theta(-k_0)\rho_{\bar{D}_0^*}(k_0,\vec{k})\theta(-q_0+k_0)\rho_\sigma^*(q_0-k_0,\vec{q}-\vec{k})\nonumber\\
&=&-\frac{1}{2}m^2G_\pi^2\int\frac{d^3k}{(2\pi)^3}\left(F(\vec{k};\Lambda)\right)^2 \nonumber\\
&&\times \frac{1}{2E_k^{\bar{D}_0^*}}\left\{\theta(q_0-E_k^{\bar{D}_0^*})\rho^*_\sigma(q_0-E_k^{\bar{D}_0^*},\vec{q}-\vec{k})+\theta(-E_k^{\bar{D}_0^*}-q_0)\rho^*_\sigma(q_0+E_k^{\bar{D}_0^*},\vec{q}-\vec{k})\right\}  \\ \nonumber\\
 {\rm Im}\tilde{\Sigma}_{\bar{D}_0^*(2c)}^{*R}(q_0,\vec{q}) &=& \frac{1}{2}\left(g_A\frac{m}{f_\pi}\right)^2\int\frac{d^4k}{(2\pi)^4}\left(F(\vec{k};\Lambda)\right)^2(q-k)^\mu (q-k)^\nu P^{\bar{D}^*}_{\mu\nu}\theta(k_0)\rho_{\bar{D}^*}(k_0,\vec{q})\theta(q_0-k_0)\rho_{\sigma}^*(q_0-k_0,\vec{q}-\vec{k}) \nonumber\\
&& - \frac{1}{2}\left(g_A\frac{m}{f_\pi}\right)^2\int\frac{d^4k}{(2\pi)^4}\left(F(\vec{k};\Lambda)\right)^2(q-k)^\mu (q-k)^\nu P^{\bar{D}^*}_{\mu\nu}\theta(-k_0)\rho_{\bar{D}^*}(k_0,\vec{q})\theta(-q_0+k_0)\rho_{\sigma}^*(q_0-k_0,\vec{q}-\vec{k}) \nonumber\\
&=&-\frac{1}{2}q_0^2\left(g_A\frac{m}{f_\pi}\right)^2\int\frac{d^3k}{(2\pi)^3}\left(F(\vec{k};\Lambda)\right)^2 \nonumber\\
&&\times\frac{1}{2E_k^{\bar{D}^*}} \frac{|\vec{k}|^2}{m_{\bar{D}^*}^2}\left\{\theta(q_0-E_k^{\bar{D}^*})\rho^*_\sigma(q_0-E_k^{\bar{D}^*},\vec{q}-\vec{k})+\theta(-E_k^{\bar{D}^*}-q_0)\rho^*_\sigma(q_0+E_k^{\bar{D}^*},\vec{q}-\vec{k})\right\}   \\ \nonumber\\ 
{\rm Im}\tilde{\Sigma}_{(2d)}^{*R}(q_0,\vec{q})  &=&\frac{3}{2}\left(g_A\frac{m}{f_\pi}\right)^2\int\frac{d^4k}{(2\pi)^4}\left(F(\vec{k};\Lambda)\right)^2(q-k)^\mu (q-k)^\nu P^{\bar{D}_1}_{\mu\nu}\theta(k_0)\rho_{\bar{D}_1}(k_0,\vec{q})\theta(q_0-k_0)\rho_{\pi}^*(q_0-k_0,\vec{q}-\vec{k}) \nonumber\\
&& - \frac{3}{2}\left(g_A\frac{m}{f_\pi}\right)^2\int\frac{d^4k}{(2\pi)^4}\left(F(\vec{k};\Lambda)\right)^2(q-k)^\mu (q-k)^\nu P^{\bar{D}_1}_{\mu\nu}\theta(-k_0)\rho_{\bar{D}_1}(k_0,\vec{q})\theta(-q_0+k_0)\rho_{\pi}^*(q_0-k_0,\vec{q}-\vec{k}) \nonumber\\
&=&-\frac{3}{2}q_0^2\left(g_A\frac{m}{f_\pi}\right)^2\int\frac{d^3k}{(2\pi)^3}\left(F(\vec{k};\Lambda)\right)^2 \nonumber\\
&&\times\frac{1}{2E_k^{\bar{D}_1}} \frac{|\vec{k}|^2}{m_{\bar{D}_1}^2}\left\{\theta(q_0-E_k^{\bar{D}_1})\rho^*_\pi(q_0-E_k^{\bar{D}_1},\vec{q}-\vec{k})+\theta(-E_k^{\bar{D}_1}-q_0)\rho^*_\pi(q_0+E_k^{\bar{D}_1},\vec{q}-\vec{k})\right\}    \ . 
\end{eqnarray}
\end{widetext}
In obtaining these expressions, we have defined the polarized tensor $P_{\mu\nu}^V$ by
\begin{eqnarray}
 P_{\mu\nu}^V \equiv g_{\mu\nu}-\frac{k_\mu k_\nu}{m_V^{*2}}
\end{eqnarray}
with $V=\bar{D}^*,\bar{D}_1$, and have used the propagator of (axial-) vector particle of the form
\begin{eqnarray}
G_{\mu\nu}(k_0,\vec{k}) = \frac{-i}{k^2-m_V^{*2}+i\epsilon}P_{\mu\nu}^V \ .
\end{eqnarray}
$\rho_\pi^*(q_0-k_0,\vec{q}-\vec{k})$ and $\rho_\sigma^*(q_0-k_0,\vec{q}-\vec{k})$ are the spectral functions for pion and $\sigma$ meson which are obtained in Eq.~(\ref{RenormalizedSpectral}), and $\rho_{X}$ ($X=\bar{D}$, $\bar{D}_0^*$, $\bar{D}^*$ and $\bar{D}_1$) are the spectral functions for ``$\bar{D}$ mesons'':
\begin{eqnarray}
\rho_{X}(k_0,\vec{k}) &=& 2\pi\epsilon(k_0)\delta(k^2-m_{X}^{*2}) \ . \label{RhoD}
\end{eqnarray}
 Note that mass of ``$\bar{D}$ mesons'' in Eq.~(\ref{RhoD}), i.e., mass of the ``$\bar{D}$ mesons'' in the internal line in self-energies in Fig.~\ref{fig:DSelfEnergy} and~\ref{fig:D0stSelfEnergy} are not observed mass in the vacuum, but modified ones in Eq~(\ref{ModMass}). Then perturbation series are defined around the mean field $\sigma_0^*$.


\end{document}